# Flexible Scheduling of Distributed Analytic Applications

Francesco Pace[1], Daniele Venzano[1], Damiano Carra[2] and Pietro Michiardi[1]

[1] Data Science Department - Eurecom, Sophia Antipolis - France, name.surname@eurecom.fr
[2] Computer Science Department - University of Verona, Verona - Italy, damiano.carra@univr.it

December 7, 2016

**Abstract**

This work addresses the problem of scheduling user-defined analytic applications, which we define as high-level compositions of frameworks, their components, and the logic necessary to carry out work. The key idea in our application definition, is to distinguish classes of components, including rigid and elastic types: the first being required for an application to make progress, the latter contributing to reduced execution times. We show that the problem of scheduling such applications poses new challenges, which existing approaches address inefficiently.

Thus, we present the design and evaluation of a novel, flexible heuristic to schedule analytic applications, that aims at high system responsiveness, by allocating resources efficiently. Our algorithm is evaluated using trace-driven simulations, with large-scale real system traces: our flexible scheduler outperforms a baseline approach across a variety of metrics, including application turnaround times, and resource allocation efficiency.

We also present the design and evaluation of a full-fledged system, which we have called Zoe, that incorporates the ideas presented in this paper, and report concrete improvements in terms of efficiency and performance, with respect to prior generations of our system.

# Contents

# 1 Introduction

The last decade has witnessed the proliferation of numerous distributed frameworks to address a variety of large-scale data analytics and processing projects. First, MapReduce [1] has been introduced to facilitate the processing of bulk data. Subsequently, more flexible tools, such as Dryad [2], Spark [3], Flink [4] and Naiad [5], to name a few, have been conceived to address the limitations and rigidity of the MapReduce programming model. Similarly, specialized libraries [6] and systems like TensorFlow [7] have seen the light to cope with large-scale machine learning problems. In addition to a fast growing ecosystem, individual frameworks are driven by a fast-pace development model, with new releases every few months, introducing substantial performance improvements. Since each framework addresses specific needs, users are left with a wide choice of tools and combination thereof, to address the various stages of their projects.

The context depicted above has driven a lot of research [8–21] in the area of resource allocation and scheduling, both from academia and the industry. These efforts materialize in cluster management systems that offer simple mechanisms for users to request the deployment of the framework they need. The general underlying idea is that of sharing cluster resources among a heterogeneous set of frameworks, as a response to static partitioning, which has been dismissed for it entails low resource utilization [8–10]. Existing systems divide the resources at different levels. Some of them, e.g. Mesos and YARN, target *low-level* orchestration of distributed computing frameworks: to this aim, they require non-trivial modifications of such frameworks to operate correctly. Others, e.g. Kubernetes [22] and Docker Swarm [23], focus on provisioning and deployment of containers, and are thus oblivious to the characteristics of the frameworks running in such containers. To the best of our knowledge, no existing tool currently addresses the problem of scheduling analytic applications as a whole, leveraging the intrinsic properties of the frameworks such applications use, but without requiring substantial modification of such frameworks.

The endeavor of this paper is to fill the gap that exists in current approaches, and *raise the level of abstraction* at which scheduling works. We introduce a general and flexible definition of *applications*, how they are composed, and how to execute them. For example, a user application addressing the training of a statistical model involves: a user-defined program implementing a learning algorithm, a framework (e.g., Spark) to execute such a program together with information about its resource requirements, the location for input and output data and possibly hyper-parameters exposed as application arguments. Users should be able to express, in a simple way, how such an application must be packaged and executed, submit it, and expect results as soon as possible.

We show that scheduling such applications represents a departure from what has been studied in the scheduling literature, and present the design of a new algorithm to address the problem. A key insight of our approach is to exploit the properties of the frameworks used by an application, and distinguish their components according to classes, core and elastic: the first being required for an application to produce work, the latter contributing to reduced execution times. Our heuristic focuses cluster resources to few applications, and uses the class of application components to pack them efficiently. Our scheduler aims at high cluster utilization and a responsive system. It can easily accommodate a variety of scheduling policies, beyond the traditional "first-come-first-served" or "processor sharing" strategies, that are currently used by most existing approaches. We study the performance of our scheduler using realistic, large-scale workload traces from Google [24, 25], and show it consistently outperforms the existing baseline approach which ignores component classes: application turnaround times are more than halved, and queuing times are drastically reduced. This induces fewer applications waiting to be served, and increases resource allocation up to 20% more than the baseline.

Finally, we present a full-fledged system, called Zoe, that schedules analytic applications according to our original algorithm and that can use sophisticated policies to determine application priorities. Our system exposes a simple and extensible configuration language that allows application definition. We validate our system with real-life experiments, and report conspicuous improvements when compared to a baseline scheduler, when using a representative workload: median turnaround times are reduced by up to 37% and median resource allocation is 20% higher.

In summary, the contributions of our work are as follows:

- We define, for the first time, a high-level construct to represent analytic applications, focusing on their heterogeneity, and their end-to-end lifecycle;
- We establish a new scheduling problem, and propose a flexible heuristic capable of handling heterogeneous requests, as well as a variety of scheduling policies, with the ultimate objective of improving system responsiveness under heavy loads;
- We evaluate our scheduling policy using realistic, large-scale workload traces and show it consistently outperforms the baseline approach;
- We build a system prototype which materializes the ideas of analytic applications and their scheduling. Our system has been in use for over one year, serving a variety of analytic application workloads. Using our new heuristic, we were able to achieve substantial improvements in terms of system responsiveness and cluster utilization.

This paper is organized as follows. We start by clarifying what analytic applications are, give examples and formulate our problem statement in Section 2. We then describe the details of our flexible scheduling heuristic, in Section 3, which we evaluate using simulations in Section 4. The system implementation is described in Section 5, and its evaluation is presented in Section 6. Finally, in Section 7 and Section 8 we discuss related work and conclude, hinting at our current research agenda.

# 2 Definitions and Problem statement

## 2.1 Definitions

We define a data analytics **framework** as a set of one or more software **components** (executable binaries) to accomplish some data processing tasks. Distributed frameworks are generally composed by a controller, a master and a number of worker components. Examples of distributed frameworks are Apache Spark [26], Google TensorFlow [27] and MPI [28]. Another example of simple data analytics framework we consider is an interactive Notebook [29].

Distributed frameworks require a *scheduler* to orchestrate their work: they execute *jobs*, each of which consists of one or more *tasks* that run in parallel the same program. Such schedulers operate at the *task level*: they assign tasks to workers, and they are highly specialized to take into account the peculiarities of each framework.

*Framework schedulers* such as Mesos [8] and Yarn [21] introduce an additional scheduling component to share cluster resources among concurrent frameworks: sharing policies are based on simple variations of Processor Sharing. Similarly, *cluster management systems* such as Docker Swarm [23] and Kubernetes [22] use a scheduler that assigns resources to generic frameworks. The problem to solve is the *efficient allocation* of resources by placing framework components and their tasks on cluster machines that satisfy a set of constraints.

We are now ready to define **analytics applications**, which are the elements we schedule in our work. Our main objective is to *raise the level of abstraction* by manipulating an abstract entity encompassing one or more analytics frameworks, their components and the necessary logic for them to cooperate toward producing useful work by running *user-defined jobs*. What sets apart our work from the state of the art is that our scheduler takes into account the notion of **component classes**, which allows modeling the specificity of each framework. We have found two distinct component classes to be sufficient to model existing analytic frameworks: thus, framework components either belong to a **core** or to an **elastic** class. Core components are compulsory for a framework to produce useful work; elastic components, instead, optionally contribute to a job, e.g. by decreasing its runtime. Consider, for example, Spark. To produce work, it needs some core components: a controller (the spark client running the DAG scheduler), a master (in a standalone deployment), and one worker (running executors). We treat additional workers as elastic components. An alternative example is an application using TensorFlow, which only works with core components: one or more parameter servers and a number of workers. These two frameworks have substantially different runtime behavior: Spark is an elastic framework that can dynamically integrate workers to dispatch tasks. TensorFlow is rigid, and uses only core components to make progress.

*To summarize, the nature of an application is that of raising the level of abstraction and an application is considered as being a collection of frameworks and their heterogeneous components as a single entity to schedule and allocate in a cluster of computers.*

## 2.2 Problem Statement

We now treat the applications defined above as abstract entities that we call *requests*: they include one or more *components*, which belong to a given class, either core or elastic. In the literature, the classical problem of scheduling generic requests to be served by a distributed system has been extensively studied [30–32]. Requests composed solely by core components are usually referred to as *rigid*, while requests composed solely by elastic components are referred to as *moldable* (if the assigned resources are decided when the request is served and they do not change for the whole execution) or *malleable* (if the resources can vary during the execution[1]). A key difference with respect to previous work is that we consider *heterogeneous requests*, composed by both core and elastic components.

For simplicity of exposition, we assume system resources that can be measured in units, and that there are $R$ available units overall to satisfy the requests. Each request $i$ specifies the amount of units for its core and elastic components, labeled $C_i$ and $E_i$ respectively. Ideally, with enough available resources, a request is allocated all of its components: in this case, we define the service (or execution) time as $T_i$. The amount of work to satisfy a request is the area of the square $W_i = T_i \times (C_i + E_i)$. More generally, a request is allocated *at least* $C_i + x_i(t)$ resources, where $0 \leq x_i(t) \leq E_i$. Then, the service time is $T_i' = \frac{W_i}{C_i + x_i(t)}$. This simple model allows updating the service time $T_i'$ when a scheduling decision modifies $x_i(t)$, by measuring the amount of work accomplished so far, and by computing the remaining amount of work to be done. While more complex models to describe $T_i'$ can be conceived, for example taking into account the multi-dimensional nature of system resources, our simple approximation doesn't affect the nature of the scheduling problem we are studying.

Essentially, the problem of scheduling the execution of an *incoming workload* of requests amounts to: *i)* sorting requests to decide in which order to serve them; *ii)* allocating distributed resources to requests selected for service. The sorting phase can be solved using naive approaches, e.g. FIFO ordering, or more sophisticated policies, that use request size information. Even more generally, requests can be placed into “pools” and be assigned priorities, to mimic the hierarchical organization of the users, for example. The allocation phase is more tricky: in the abstract, it is a “packing” problem that has to decide how to shape requests being served. Even assuming service times to be known a-priori (e.g., $T_i$ is given as an input), it is well known that the *on-line scheduling problem* is NP-hard [30]. Therefore, we need to find a suitable heuristic to approximate a solution to the scheduling optimization problem. In our case, it amounts to minimizing the application *turnaround times*, which is the interval of time between request $i$ submission and its completion. In the context we consider, optimizing the average turnaround time represents a meaningful performance metric, as it caters system responsiveness. Next, we motivate our problem with a simple illustrative example.
**Illustrative example.** We consider a system with 10 available resource units, and four requests waiting to be served, as shown in Figure 1. Each request needs 3 units for the core components, and different units for the elastic components. For each request, $T_i = 10$.

[1]An example of *malleable* framework is Spark [33]. Worker can be added or removed without destroying the application execution.

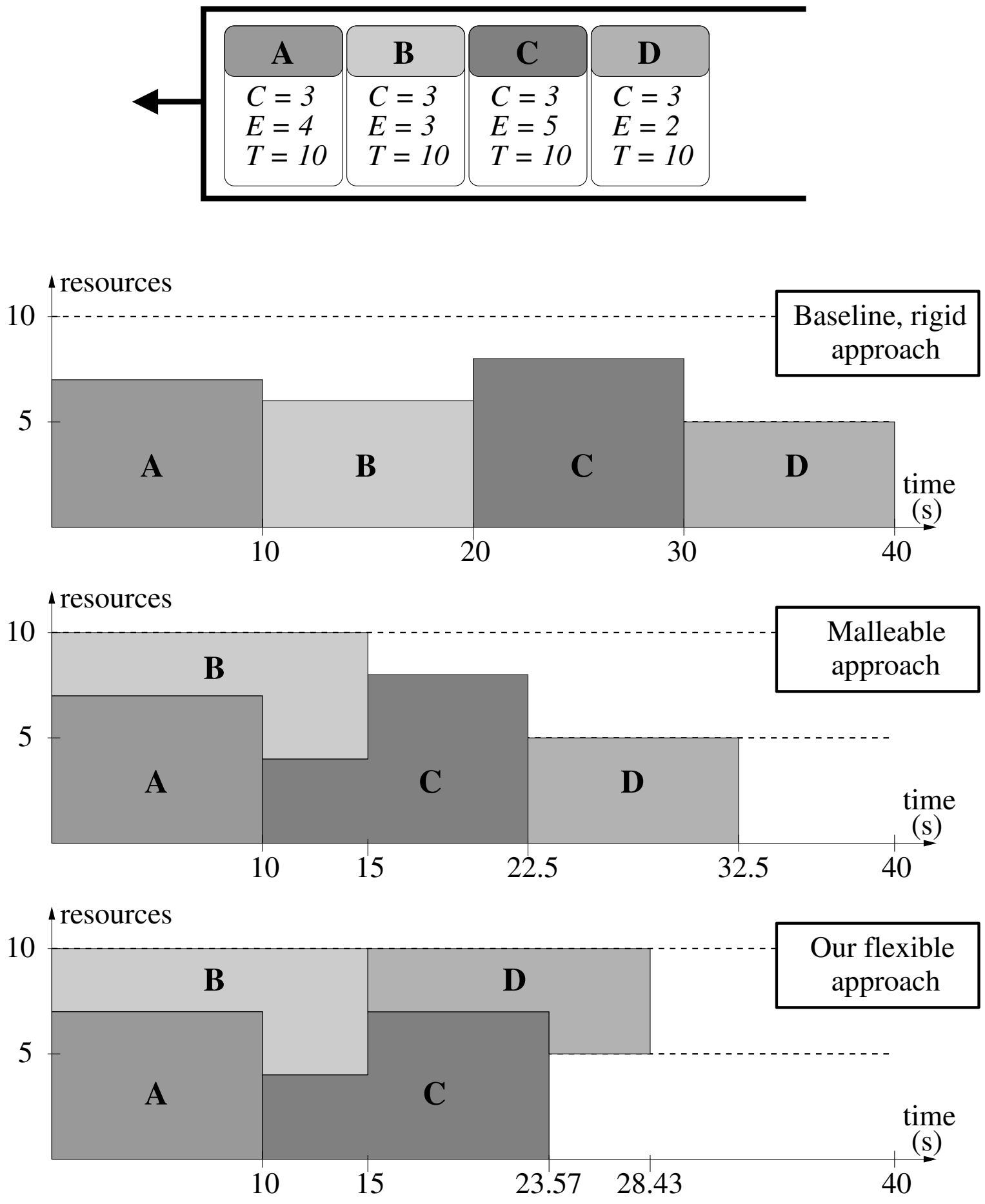

Figure 1: Illustrative examples of request scheduling: (top) rigid, (middle) malleable, (bottom) flexible approaches.

Given these requests, a traditional, rigid approach to scheduling – which does not make the distinction between component classes – assigns all required resources to each request. Since all requests need at least 5 units ($C_i + E_i \geq 5$), and since any pair of requests have an aggregated need that exceeds 10 units, the scheduler serves one request at a time (Figure 1, top): the average turnaround time is 25s. Note that, in this case, backfilling is not possible, i.e., even by changing the order in which requests are served the situation does not change.

Another scheduling approach comes from the literature of malleable job scheduling. The scheduler assigns all resources to the first request in the waiting line, then assigns the remaining resources (if any) to the next request, and so on, until no more free resources are available. This heuristic has been shown to be close to optimal [31]. Figure 1, middle, illustrates the idea: request B can be served along with request A. When request A has completed, the scheduler first assigns more resources to request B, and then tries to serve the next request. Similarly, when request B has completed, the scheduler first assigns more resources to request C, then attempts at serving request D. However, since request D needs at least $C_i = 3$ units, the scheduler is blocked (note that request C uses 8 units), so request D needs to wait, and some system resources remain unused. The average turnaround time is 20s.

In this work we advocate the need for a new approach to scheduling, which distinguishes component classes. The idea is to exploit the flexibility of elastic components and use system resources more efficiently. Intuitively, a solution to the problems of existing heuristics is to reclaim some resources assigned to elastic components of a running request and assign them to a pending request. This is shown in the bottom of Figure 1: the scheduler reclaims just one unit from request C so that it can provide 3 units to request D, which are sufficient for starting its core components and produce useful work. With this approach, the average turnaround is 19.25s.

While the above solution seems simple, it poses many challenges: how many units assigned to elastic components can be sacrificed for serving the next request? How many requests should be served concurrently? Should the scheduler focus on core components alone, to make sure many requests are served concurrently? How can scheduling take into account the priorities assigned by the sorting phase?

The last point introduces an additional challenge, related to *preemptive scheduling* policies. If a high priority request arrives, since it is not possible to interrupt core components – for this would kill the request – how can we select and preempt elastic components to accommodate the new request?

*Given heterogeneous, composite requests, which are neither rigid, nor malleable (but both), available scheduling heuristics in the literature fall short in addressing the sorting and allocation problems: a new approach is thus truly desirable.*

# 3 A Flexible Scheduling Algorithm

## 3.1 Design guidelines

We characterize a request by its arrival time, its priority (to decide the order in which the requests should be served), the resources it asks for (core and elastic) and the execution time (in isolation, *i.e.,* when all required resources are granted to the application). Given an incoming workload, our goal is to optimize the sum of the turnaround times $\tau_i$, that is:

$$\min \sum_i \tau_i \Rightarrow \min \sum_i (\text{queuing}_i + \text{execution}_i)$$

The actual execution time depends on the amount of resources assigned over time to the request. Now, recall that the scheduling problem can be broken into sorting and allocation phases. Sorting determines when a request is served, thus it has an impact on its queuing time. The allocation phase contributes both to queuing and actual execution times. Depending on allocation granularity [9], a request might need to wait for a number of resources to be available before occupying them, thus increasing – albeit indirectly – the queuing time. The execution time is directly related to the allocation algorithm and to the workload characteristics.

In this work we decouple request sorting from allocation:[2] our scheduler maintains the request ordering, as imposed by an external component, and only focuses on resource allocation. Sorting can be simply based on arrival times (which amounts to implement a FIFO queuing discipline), or can use additional information, such as request size (thus implementing a variety of size-based disciplines).

Overall, we optimize request turnaround times through careful resource allocation, and *design an algorithm that strives at allocating all available cluster resources, by serving the least number of requests at a time.* Intuitively, by "focusing" resources to few requests, we expect their execution times to be small. Consequently, queued requests also enjoy smaller wait times, because resources are freed more quickly.

## 3.2 Algorithm Details

Although we support preemptive scheduling policies, to simplify exposition, we first consider the case with no preemption: resources assigned to a request can only increase, and a new request can be placed, at most, at the head of the waiting line, depending on the sorting component. We stress that the output of our scheduling algorithm is a *virtual assignment*, *i.e.,* the mechanism to physically allocate resources according to the computed assignment (core and elastic components for running applications) is separate from the scheduling logic, and considered as an implementation detail.

Our resource allocation procedure is called REBALANCE, and it is triggered by two events: request arrivals and departures – see Algorithm 1, ignoring highlighted lines. When a new request arrives (procedure ONREQUESTARRIVAL), the resource assignment is done only if such a request is placed at the head of the waiting line and there are unused resources that are sufficient for running its core components. When a request is completed (procedure ONREQUESTDEPARTURE), the released resources are always reassigned.

[2]This approach is similar to the one used in the SLURM scheduler [34], where the order of the pending jobs is given by an external, pluggable, component, and the scheduler processes the jobs following that order.

**Algorithm 1:** Resource assignment procedures

```
procedure OnRequestArrival(req)
    if req.P > S.tail.P then
        if req.C ≤ Σ_{j∈S} req_j.E then
            Insert(req, S)
            Rebalance( )
        else
            Insert(req, W)
    else
        Insert(req, L)
        if req == L.head and req.C ≤ avail then
            Rebalance( )

procedure OnRequestDeparture( )
    while W.head.C + Σ_{j∈S} req_j.C < total and (W not ∅) do
        Insert(pop(W), S)
    Rebalance( )

procedure Rebalance( )
    while Σ_{j∈S} (req_j.C + req_j.E) < total and (L not ∅) do
        req ← L.head
        if req.C + Σ_{j∈S} req_j.C < total then
            Insert(pop(L), S)
        else
            break
    avail ← total − Σ_{j∈S} req_j.C
    forall req ∈ S do
        req.G ← 0
    req ← S.head
    while avail > 0 and (req not null) do
        req.G ← min(req.E, avail)
        avail ← (avail − req.G)
        req ← req.next
```

The scheduler maintains two ordered sets: the requests waiting to be served ($\mathcal{L}$), and the requests in service ($\mathcal{S}$). Each request $req$ needs $req.C$ core components and $req.E$ elastic components; depending on the allocation, request $req$ is granted $0 \leq req.G \leq req.E$ elastic components. The core of the procedure REBALANCE (lines 27-30) operates as follows: each request $req$ in the serving set $\mathcal{S}$ has always **at least** $req.C$ resources assigned. Excess resources are assigned to the requests in $\mathcal{S}$ following the request order. The scheduler assigns as many elastic components as possible to the first request, then to the second, and so on, in cascade.

Following the design guidelines, the set $\mathcal{S}$ should only contain the requests that are strictly necessary to use all the available resources. This is accomplished by the first part of the procedure REBALANCE (lines 17-22): a request is added to $\mathcal{S}$ if the current requests in $\mathcal{S}$ are not able to saturate the total resources (*total*, line 17). Note that we add a request to $\mathcal{S}$ only if there is room to allocate all of its core components.

### 3.3 Preemptive policies

We now consider preemptive policies: request arrivals can trigger (partial) preemption of running requests, e.g. if new requests have higher priority than that of the last request in service. In this case, the tuple describing a request also stores its priority, $req.P$. It is important to note that, in this work, the preemption mechanism only operates on elastic components of running applications, whereas core components (that are vital for an application) cannot be preempted.

The highlighted lines in Algorithm 1 show the modifications to the procedures ONREQUESTARRIVAL and ONREQUESTDEPARTURE to support preemption. When a new request arrives, if its priority is higher than the requests in service, we check if its core components can be allocated using the resources occupied by the elastic components of currently running requests. If so, we insert the request into the set $\mathcal{S}$ and call REBALANCE. Otherwise, we insert the request into an auxiliary waiting line $\mathcal{W}$, which is given priority when resources become available. Indeed, procedure ONREQUESTDEPARTURE indicates that we first consider the waiting requests in $\mathcal{W}$, and we add to the set $\mathcal{S}$ as many of them as possible, considering solely the core components. In other words, requests in $\mathcal{W}$ have higher priority than those in $\mathcal{L}$. Finally, the call of REBALANCE assigns the remaining resources to the elastic components of high priority requests.

## 4 Numerical evaluation

### 4.1 Methodology

We evaluate our algorithm using an event-based, trace-driven discrete simulator developed to study the scheduler Omega [9], which we extended in order to make it work with applications, instead of low-level jobs and to use the concept of component classes. Our scheduler implementation supports a variety of policies[3]: we present results for the FIFO and the shortest job first (SJF) policies, which further optimizes system responsiveness. Our implementation first obtains a "virtual assignment" with Algorithm 1, then fulfills it by allocating resources accordingly, which happens instantaneously. Additionally, we have implemented a baseline, consisting of a rigid scheduler that does not distinguish component classes, which is representative of current cluster management systems. In our simulations, we consider two-dimensional resources, including definitions of CPU and RAM requirements. We would like to stress that the "virtual assignment" can take into consideration other constraints as well (e.g., GPU).

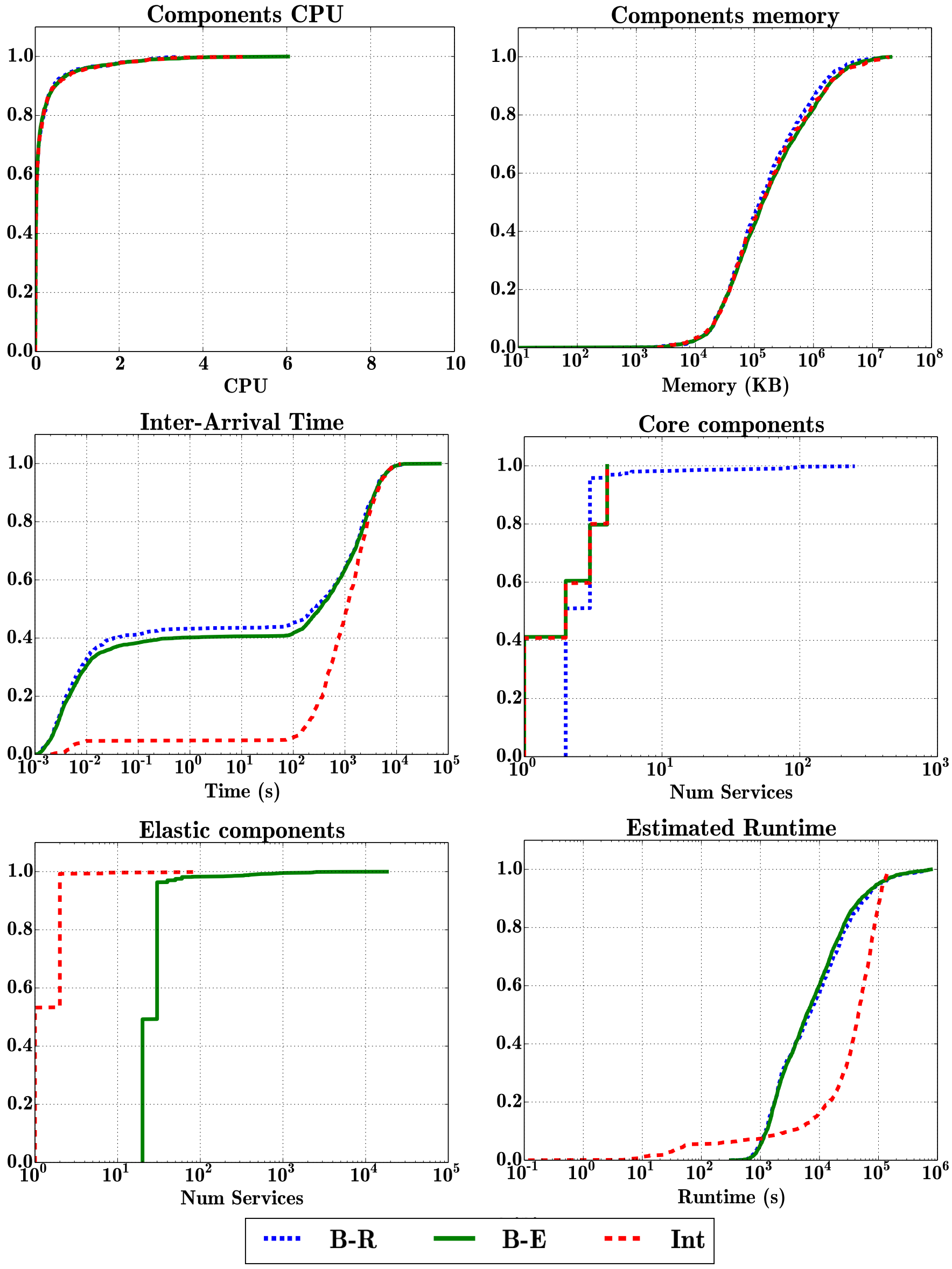

Figure 2: Workload Definition.

Our scheduler currently accepts **application workloads** of two kinds. The first is **batch applications**, that take from a few seconds to a few days to complete: these are delay-tolerant applications, with a very simple life-cycle. Core components must first start to produce useful work, by executing user-defined jobs that are "passed" to the application; eventually, elastic components can contribute to the application progress. Once the user programs are concluded, the application finishes, releasing resources occupied by its frameworks and components. The second is **interactive applications**, which involve a "human in the loop": these are latency-sensitive applications, with a life-cycle triggered by human activity. In this case, core components must start as soon as possible, to allow user interaction with the application (e.g., a Notebook).

For our performance evaluation, we use publicly available traces [24,25], and generate a workload by sampling the empirical distributions we compute from such traces. First, we focus on batch applications alone, and simulate both *rigid* (e.g. TensorFlow) and *elastic* (e.g. Spark) variants: the label **B-R** represents rigid applications with only core components; the label **B-E** stands for elastic applications, with both core and elastic components. Then, we evaluate the benefit of preemption by complementing the above workload with (simulated) interactive applications.

Figure 2 shows the definition of the workload: in particular we can see the CDFs for requested CPU and memory, then the inter-arrival time and estimated run time and finally, the number of core and elastic components. Batch applications are assigned a number ranging from a few to tens of thousands of components. Instead, interactive applications are smaller, and use up to hundreds of elastic components. The resource requirements of application components follow that of the input traces, ranging from few MB to a few dozens GB of memory, and up to 6 cores. Application *runtime* is generated according to the input traces, and range from a few dozen seconds to several weeks (of simulated time). Application inter-arrival times are drawn from the empirical distributions of the input traces, and exhibit a bi-modal distribution with fast-paced bursts, as well as longer intervals between application submissions. In summary, our workload consists of 80,000 applications, with 80% batch and 20% interactive applications. Batch applications include 80% elastic and 20%

[3]We support size-based disciplines in the family of SMART policies [35]. Here we assume application size information to be known a-priori.

rigid variants.

We simulate a cluster consisting of 100 machines, each with 32 cores and 128GB of memory. All results shown here include 10 simulation runs, for a total of roughly 3 months of simulation time for each run.

Finally, the metrics we use to analyze the results include: application **turnaround**, which allow to reason about the scheduling objective function, and **queuing time**, which is an important factor contributing to the turnaround time. Additionally, we measure the **queue sizes** that hold pending and running applications, and **resource allocation**, measured as the percentage of CPU and memory the scheduler allocates to each application.

## 4.2 Comparison with the baseline

We now perform a comparative analysis between our flexible scheduler and the baseline, and start by disabling preemption: we omit interactive applications from the workload. Figure 3 (left) illustrates the most important percentiles (in a box-plot) of the distribution of turnaround times, where we compare two policies (namely, FIFO and SJF) used by both the baseline and our scheduler. The benefits of our approach are noticeable, irrespectively of the scheduling discipline: the median turnaround is halved when compared to the baseline, indicating superior system responsiveness. Additionally, we observe the benefits of a size-based policy in further decreasing turnaround times. We note that our approach is beneficial for both rigid and elastic batch applications: Figure 3 (center) shows a box-plot of application queuing times, which contribute to their turnaround. With our approach, both kinds of applications spend less time waiting in a queue to be served. By differentiating classes of components, applications can execute as soon as enough resources to produce work are available. Finally, Figure 3 (right) focuses on application runtime: we report the **slowdown** computed as the ratio between the nominal application runtime (*i.e.*, the time required for an application to complete in an empty system, with all application components allocated their requested resources) and the effective application runtime obtained with the simulation. Values above one indicate that applications run slower in a system absorbing a given workload when compared to applications running in an empty system. Overall, these results show that our scheduling approach does not impose a high toll on application runtime, while globally contributing to improved turnaround times.

Next, we support the general results discussed above with additional details. Figure 4 shows the box-plots of the distribution of queue sizes, for both the pending and the running queues. Our approach induces a smaller number of applications waiting to be served, as well as a larger number of applications running in the system, compared to the baseline and across different policies. Indeed, our flexible scheduler achieves a better packing of applications, which means they can start sooner. Additionally, the benefits of a size-based discipline are clear: the number of applications waiting is almost one order of magnitude smaller compared to a FIFO policy, while the number of running applications is similar.

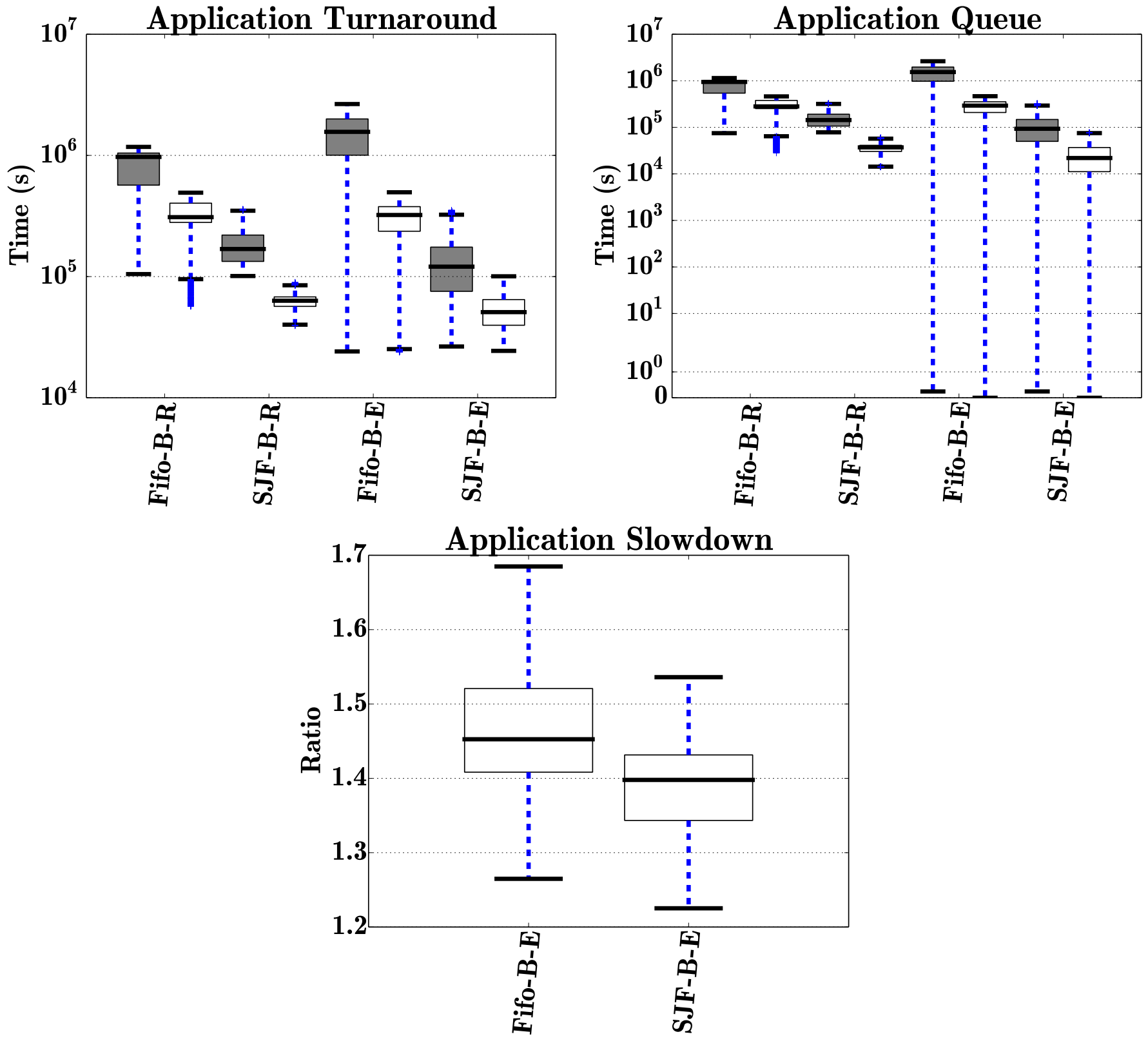

Figure 3: Comparison of turnaround and queue time distributions, and application slowdown distributions for FIFO and SJF policies. White boxes (right box of every pair) corresponds to our flexible scheduler, gray boxes correspond to the baseline. *B-E* stands for batch elastic and *B-R* stands for batch rigid applications.

Figure 5 shows metrics from the cluster perspective: our approach (for both disciplines) induces a far better resource allocation compared to the baseline, achieving more than 20% gains in both CPU and RAM allocation.[4]

Finally, we are going to show more results, that compare a rigid, a malleable and our flexible scheduler with different policies. It is worth mentioning that currently no solution support a malleable scheduler as we presented it in Section 2.2.

In single-server systems, the policies that are optimal for minimizing the average turn around time are called SMART [35]; they prioritize short application over longer one. Two example of SMART policies are SFJ (Shorted-Job First) and SRPT (Shortest-Remaining-Processing-Time). However, even if SRPT is considered optimal, it is rarely used because it leads to starvation of long running application. For this reason policies like HRRN (Higher-Response-Ration Next) can be used; they calculate a virtual size that is updated the longer the application reside in queue. In this evaluation we are going to use four different policies: FIFO, SFJ, SRPT and HRRN.

[4]Allocation is different from utilization: the simulator does not account for real application execution, so we cannot report utilization figures.

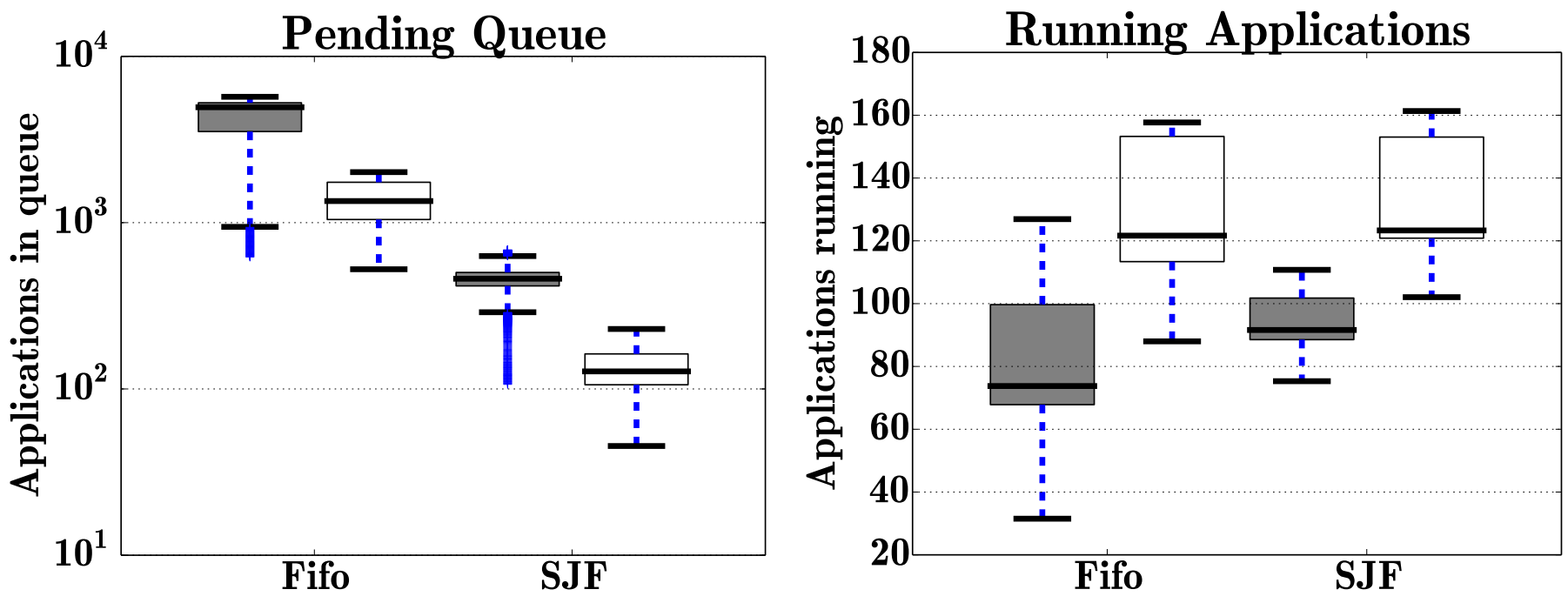

Figure 4: Comparison of queues size for FIFO and SJF between our flexible scheduler and the baseline. The white boxes (right box of every group) correspond to our flexible algorithm, gray boxes to the baseline.

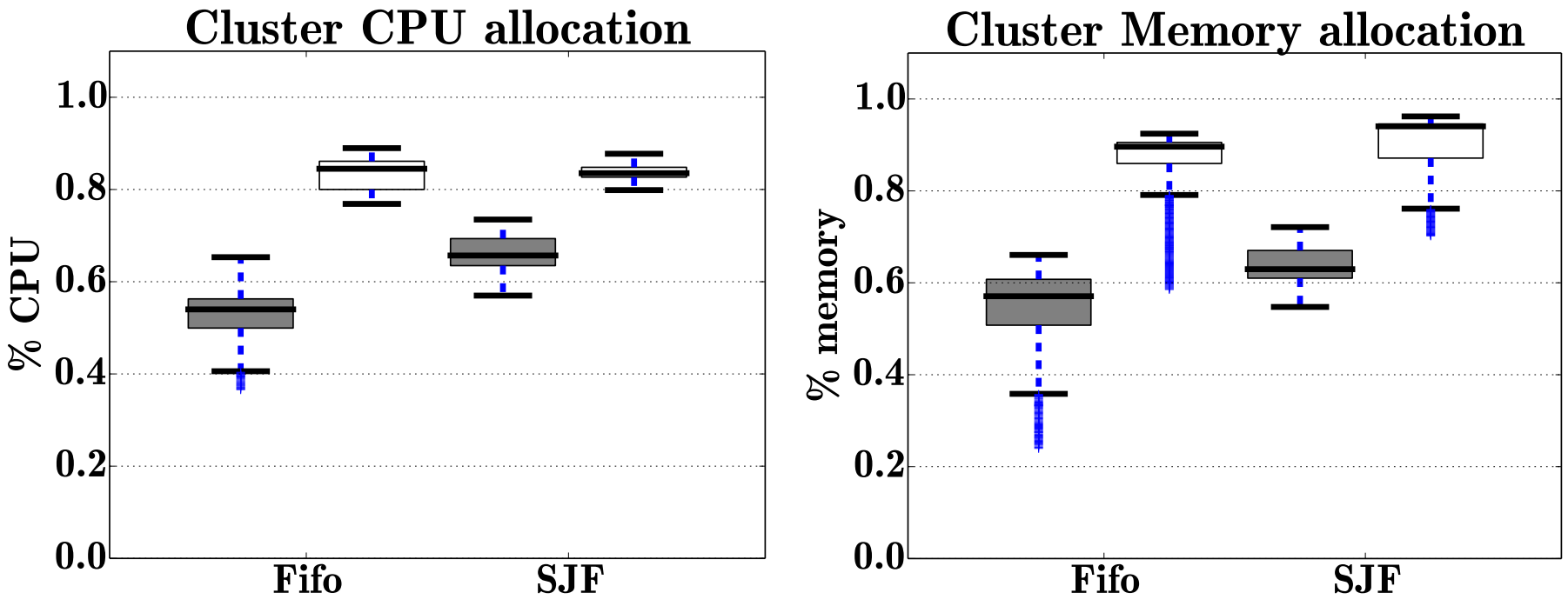

Figure 5: Comparison of resource allocation distributions for FIFO and SJF policies, between our flexible scheduler and the baseline. White boxes (right box of every pair) correspond to our approach, dashed boxes to the baseline.

Figures 6 to 13 confirm that our solution performs far better than a rigid scheduler and slightly better than a malleable regarding of the policy used to sort the queue.

*In conclusion using our flexible scheduler can greatly reduce the turn around time while improving resources allocation regardless of the policy that is used to sort the applications in the pending queue.*

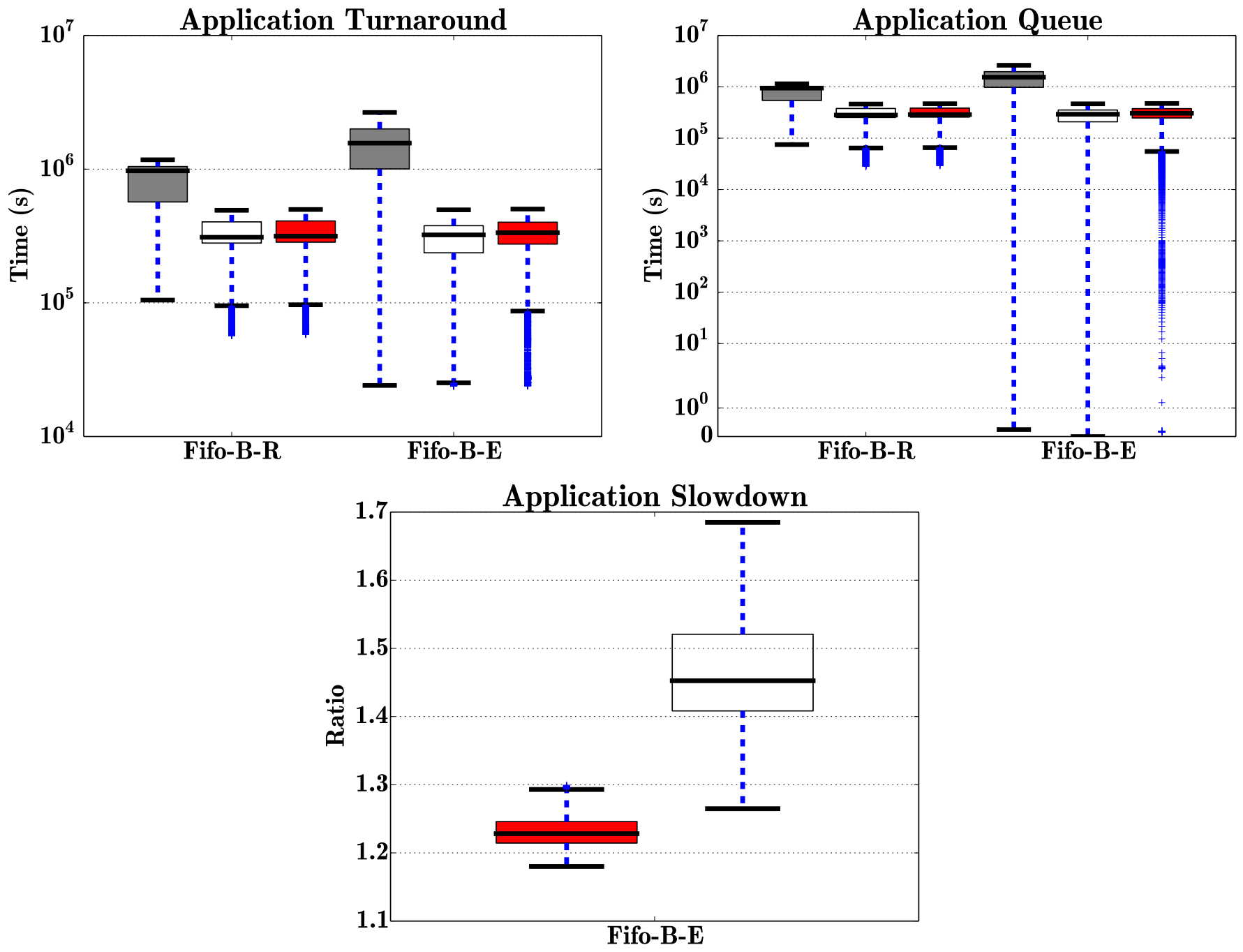

Figure 6: Comparison using FIFO between the rigid, the elastic and our system. For every group, the boxes in the middle correspond to our system, the boxes on the left correspond to the rigid system while the boxes on the right are for the elastic system. *B-E* stands for batch elastic and *B-R* stands for batch rigid applications.

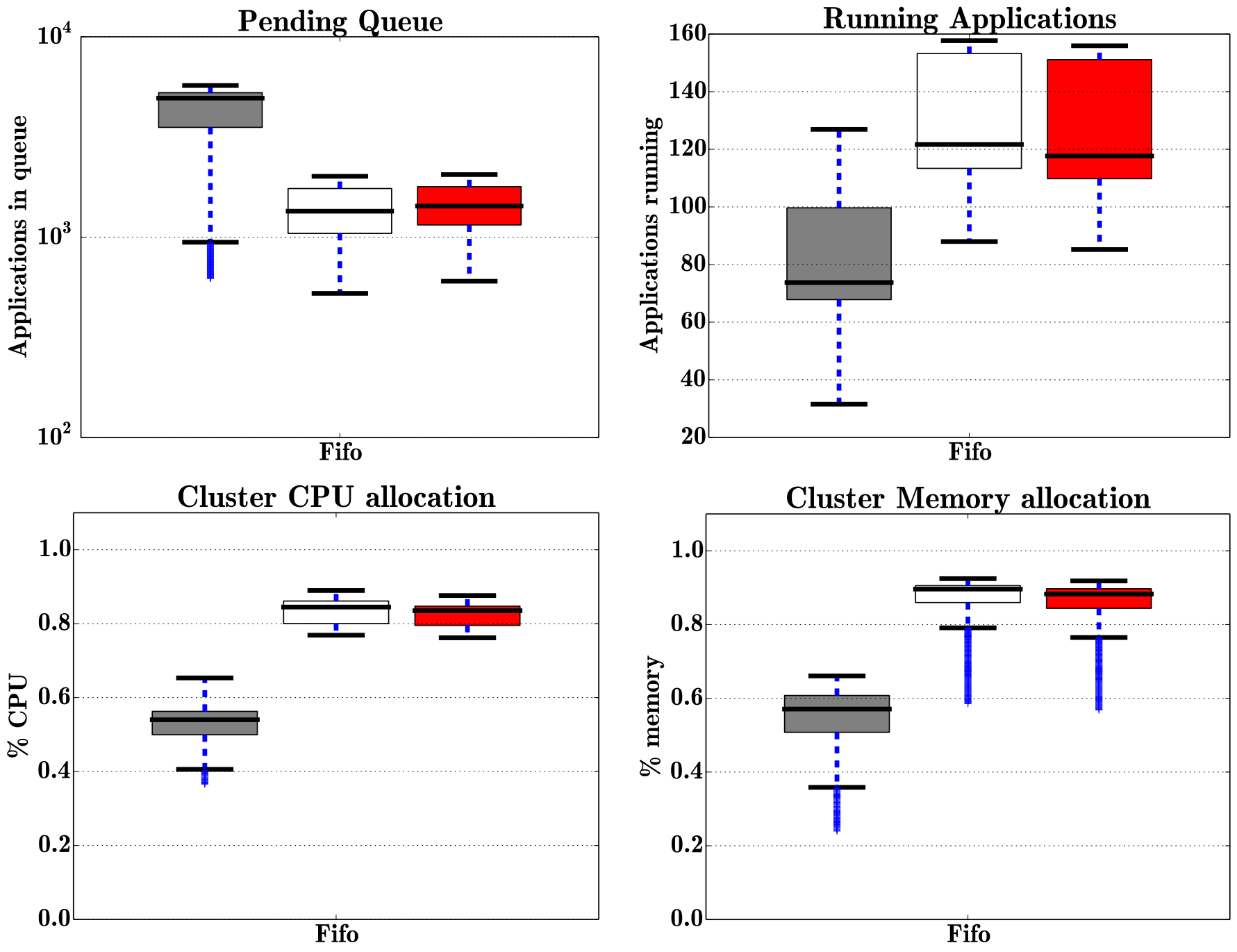

Figure 7: Comparison using FIFO between the rigid, the elastic and our system. For every group, the boxes in the middle correspond to our system, the boxes on the left correspond to the rigid system while the boxes on the right are for the elastic system. *B-E* stands for batch elastic and *B-R* stands for batch rigid applications.

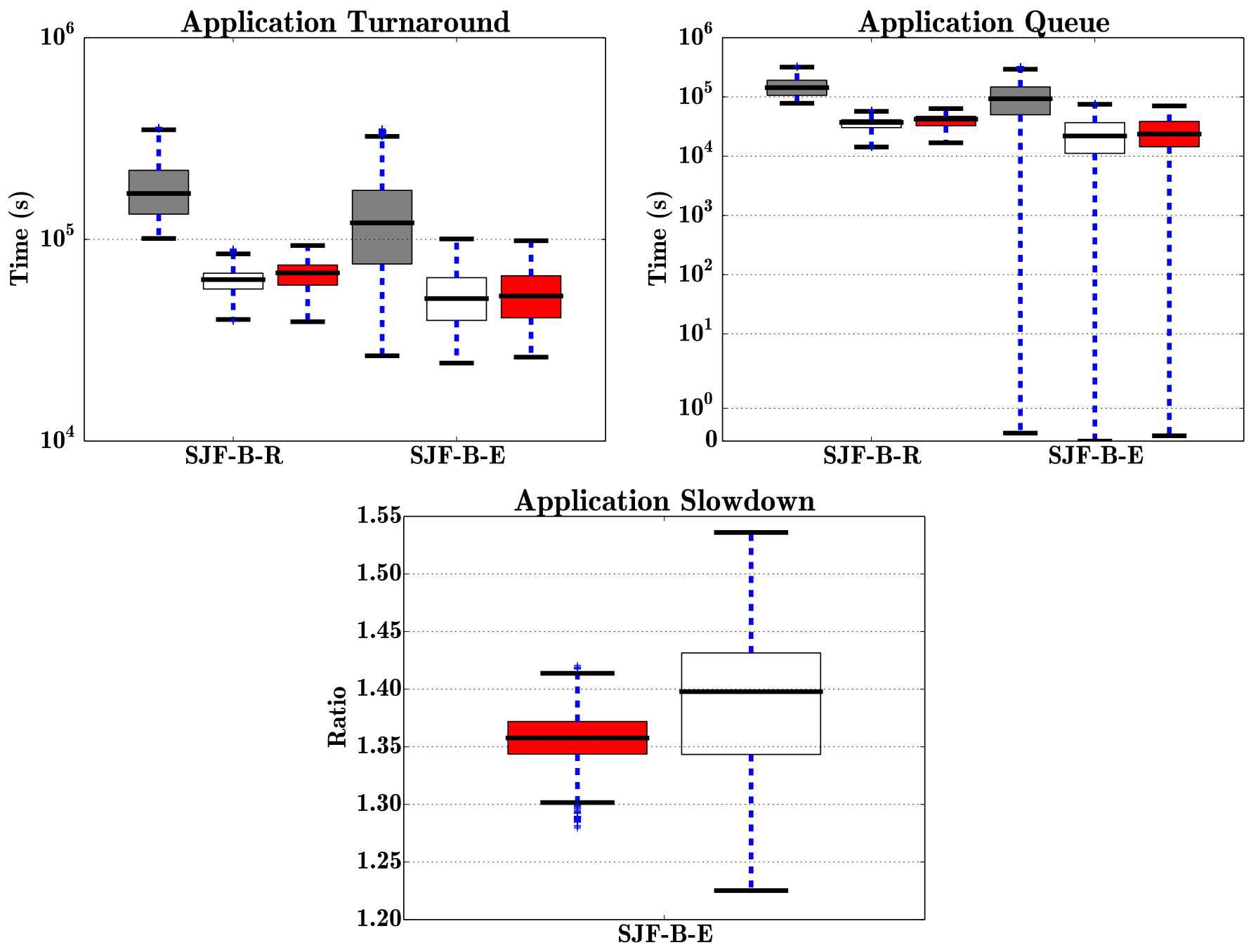

Figure 8: Comparison using SJF between the rigid, the elastic and our system. For every group, the boxes in the middle correspond to our system, the boxes on the left correspond to the rigid system while the boxes on the right are for the elastic system. *B-E* stands for batch elastic and *B-R* stands for batch rigid applications.

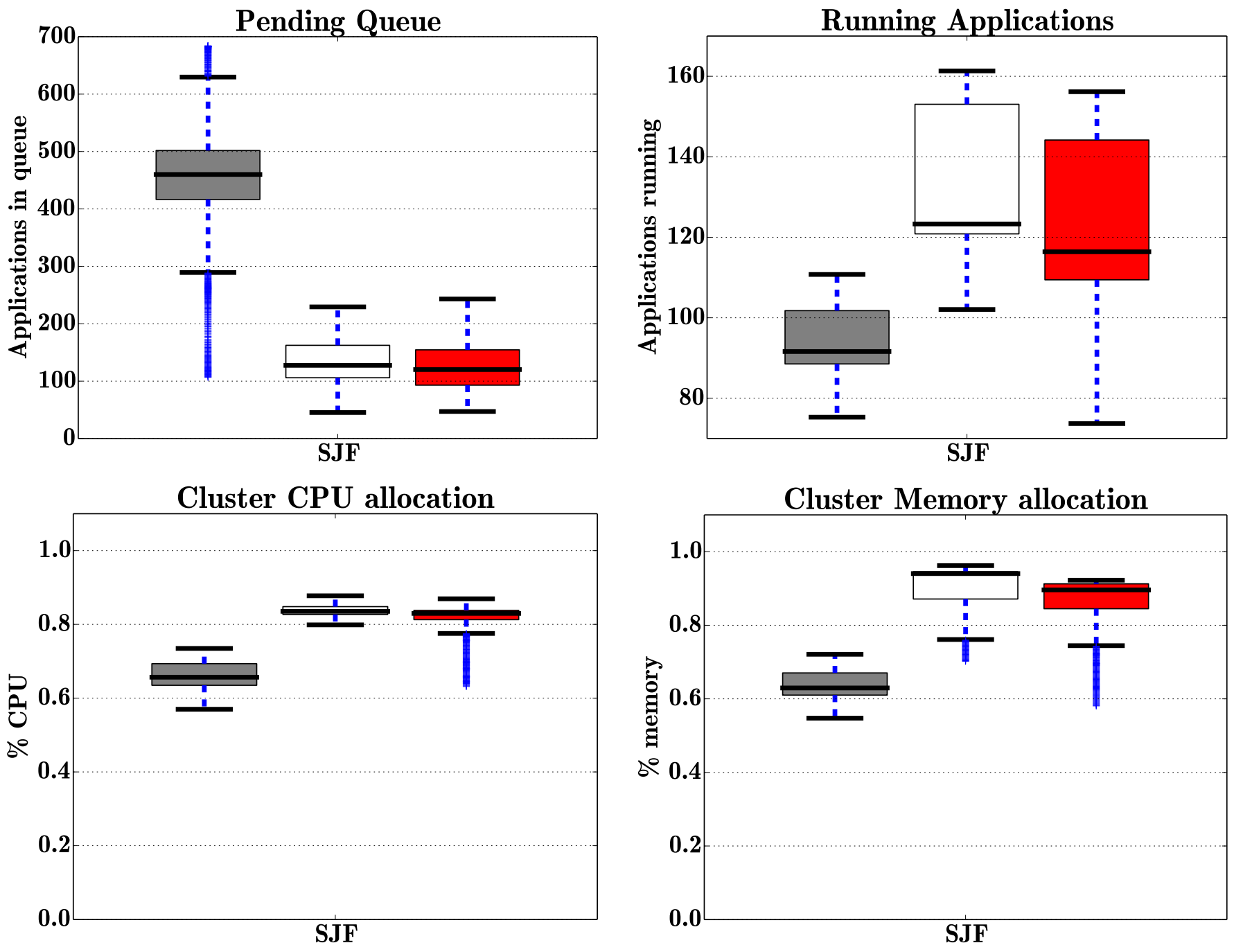

Figure 9: Comparison using SJF between the rigid, the elastic and our system. For every group, the boxes in the middle correspond to our system, the boxes on the left correspond to the rigid system while the boxes on the right are for the elastic system. *B-E* stands for batch elastic and *B-R* stands for batch rigid applications.

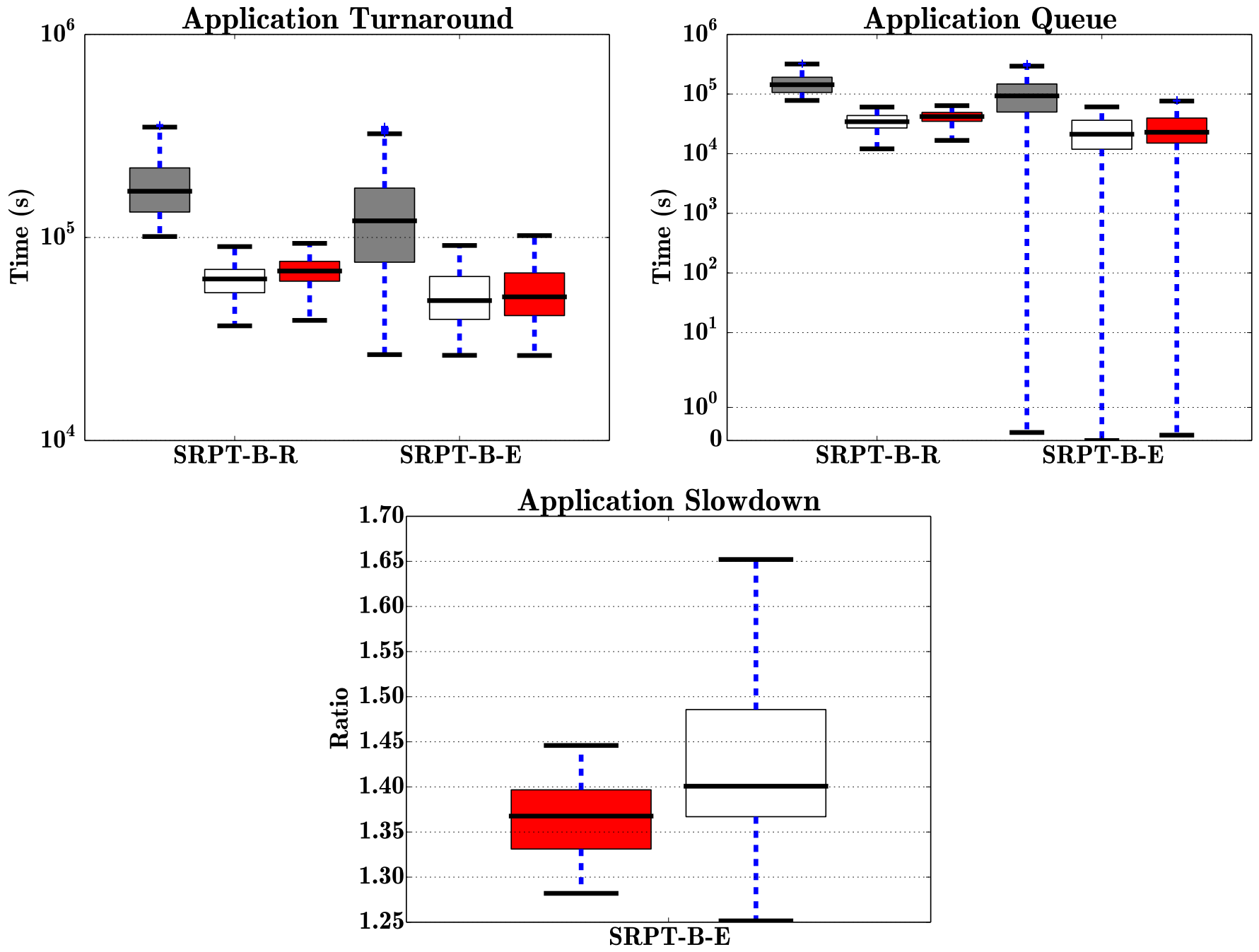

Figure 10: Comparison using SRPT between the rigid, the elastic and our system. For every group, the boxes in the middle correspond to our system, the boxes on the left correspond to the rigid system while the boxes on the right are for the elastic system. *B-E* stands for batch elastic and *B-R* stands for batch rigid applications.

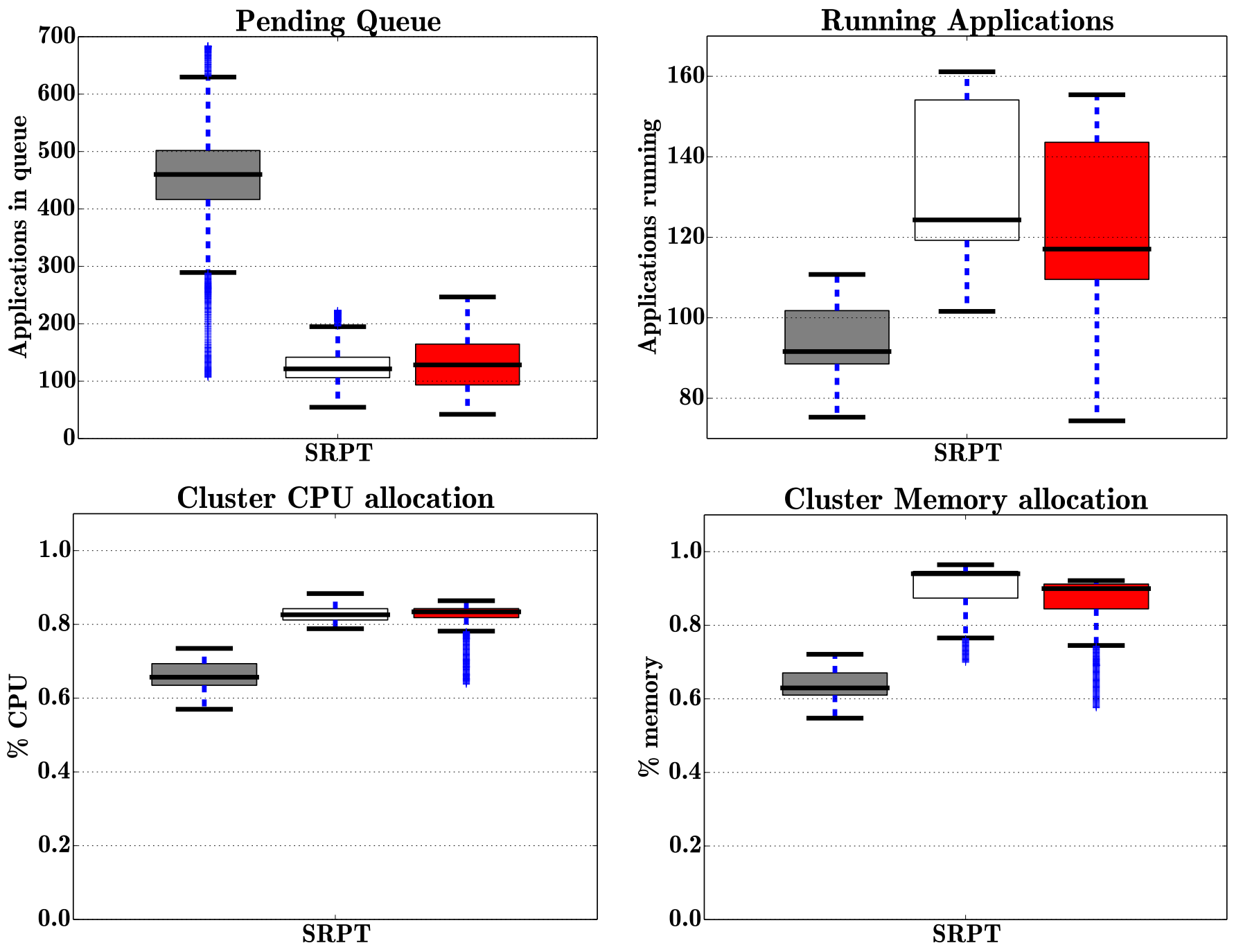

Figure 11: Comparison using SRPT between the rigid, the elastic and our system. For every group, the boxes in the middle correspond to our system, the boxes on the left correspond to the rigid system while the boxes on the right are for the elastic system. *B-E* stands for batch elastic and *B-R* stands for batch rigid applications.

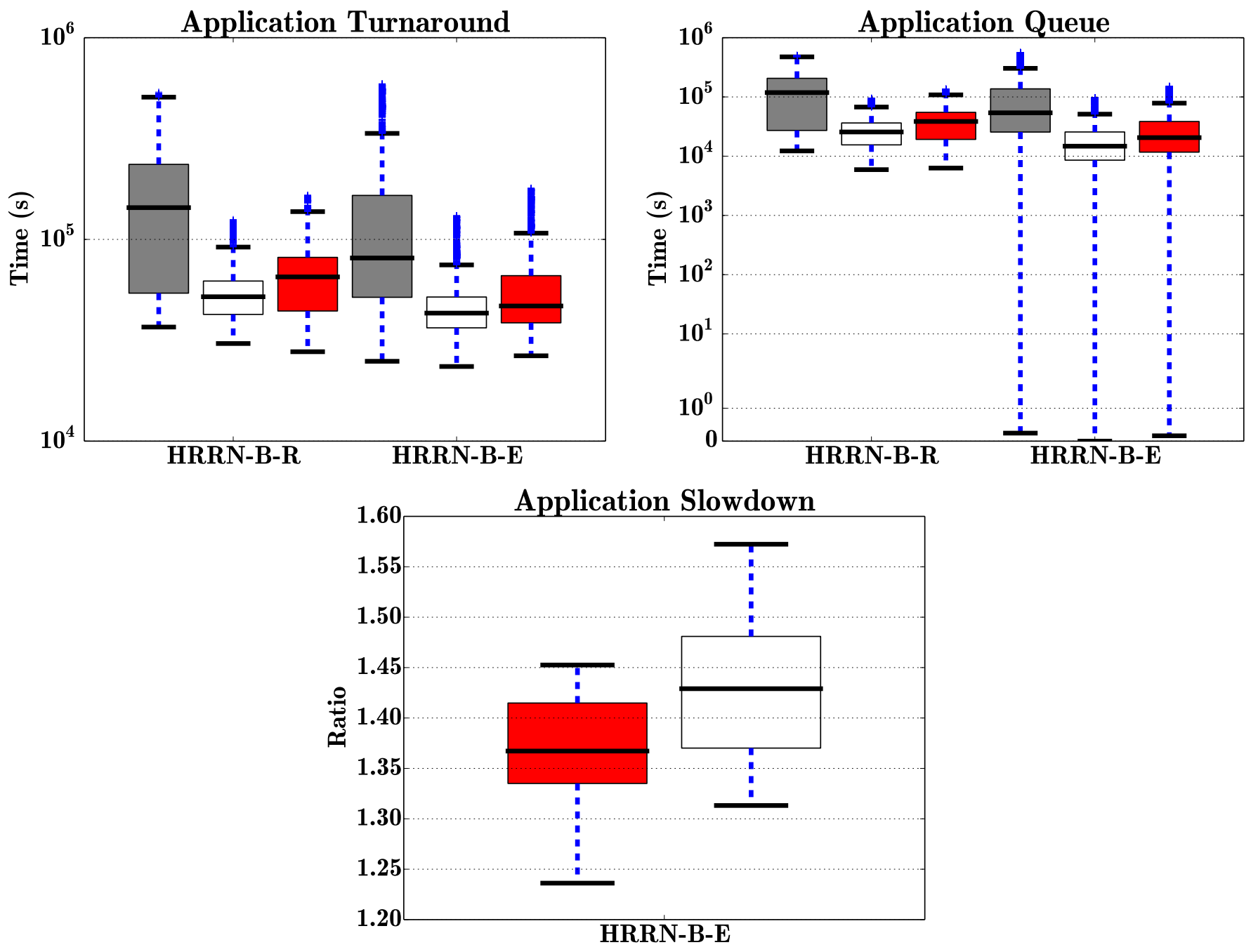

Figure 12: Comparison using HRRN between the rigid, the elastic and our system. For every group, the boxes in the middle correspond to our system, the boxes on the left correspond to the rigid system while the boxes on the right are for the elastic system. *B-E* stands for batch elastic and *B-R* stands for batch rigid applications.

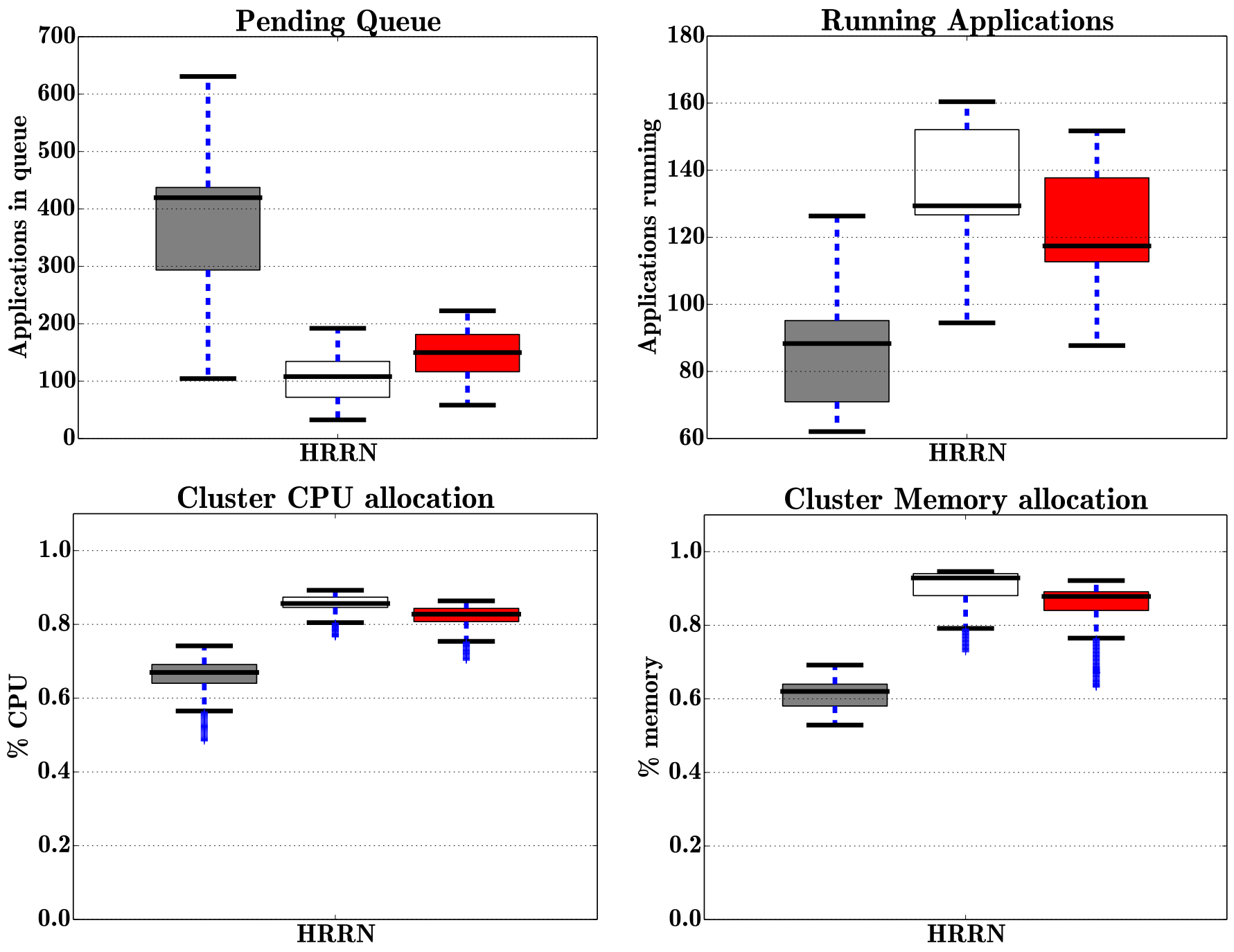

Figure 13: Comparison using HRRN between the rigid, the elastic and our system. For every group, the boxes in the middle correspond to our system, the boxes on the left correspond to the rigid system while the boxes on the right are for the elastic system. *B-E* stands for batch elastic and *B-R* stands for batch rigid applications.

| Name | Definition |
|---|---|
| SJF-2D | $runTime * \#RequestedServices$ |
| SRPT-2D1 | $remainingRunTime * \#RequestedServices$ |
| SRPT-2D2 | $remainingRunTime * \#ServicesYetToBeScheduled$ |
| HRRN-2D | $(1 + \frac{waitTime}{runTime}) * \#RequestedServices$ |
| SJF-3D | $runTime * \sum_{i=1}^{services} CPU_i * RAM_i$ |
| SRPT-3D1 | $remainingRunTime * \sum_{i=1}^{services} CPU_i * RAM_i$ |
| SRPT-3D2 | $remainingRunTime * \sum_{i=1}^{servicesToSchedule} CPU_i * RAM_i$ |
| HRRN-3D | $(1 + \frac{waitTime}{runTime}) * \sum_{i=1}^{services} CPU_i * RAM_i$ |

Table 1: Definition of size used in the evaluation

## 4.3 Comparison between different definitions of size

In this section we will focus on size-based policies and in particular on their definition of size; they were meant to be used in a single-server system or an unidimensional world, where there is only one type of resource. In a data center, however, there is a cluster of machines, and each one has at least two type of resources: CPU and Memory. For this reason we compare different definitions of size that will move from an unidimensional to a multidimensional world; similarly to [36], we want to "improve" the definition of size for every single policies that we introduced in the previous section by, progressively, adding more information to them. Table 1 shows the definition of size that we use in this section of the evaluation.

| SJF-2D | SRPT-2D1 | SRPT-2D2 | HRRN-2D | SJF-3D | SRPT-3D1 | SRPT-3D2 | HRRN-3D |
|---|---|---|---|---|---|---|---|
| 38,340.68 | 38,767.29 | 39,425.82 | 89,596.90 | 33,451.60 | 33,418.51 | 33,413.45 | 229,691.64 |

Table 2: Turn around time (in second) between different definitions of size in our flexible scheduler.

Table 2 shows the results obtained by running the previous workload with these definitions of size. We see that there is an improvement when we add more information to the definition of size. The only policy that perform worst in all cases is HRRN; the results are expected since the policy avoid starvation by increasing the priority of applications while they remain in the queue, thus big and long running applications can start before short ones. Figures 14 to 28 show the difference between all the size definitions that we took in exam under a rigid, malleable and our flexible scheduler.

These results are possible because, in a unidimensional world, a short-living application can block the queue due to the high amount of resources requested. To avoid this situation we can add the information of the number of services required to actually allow small applications to be executed before large ones; as we see in Figures 17 to 28, this information improves the turn around time in all policies but HRRN. We would expect that adding more information to the size could further improve the turn around time, instead we see that this is not true

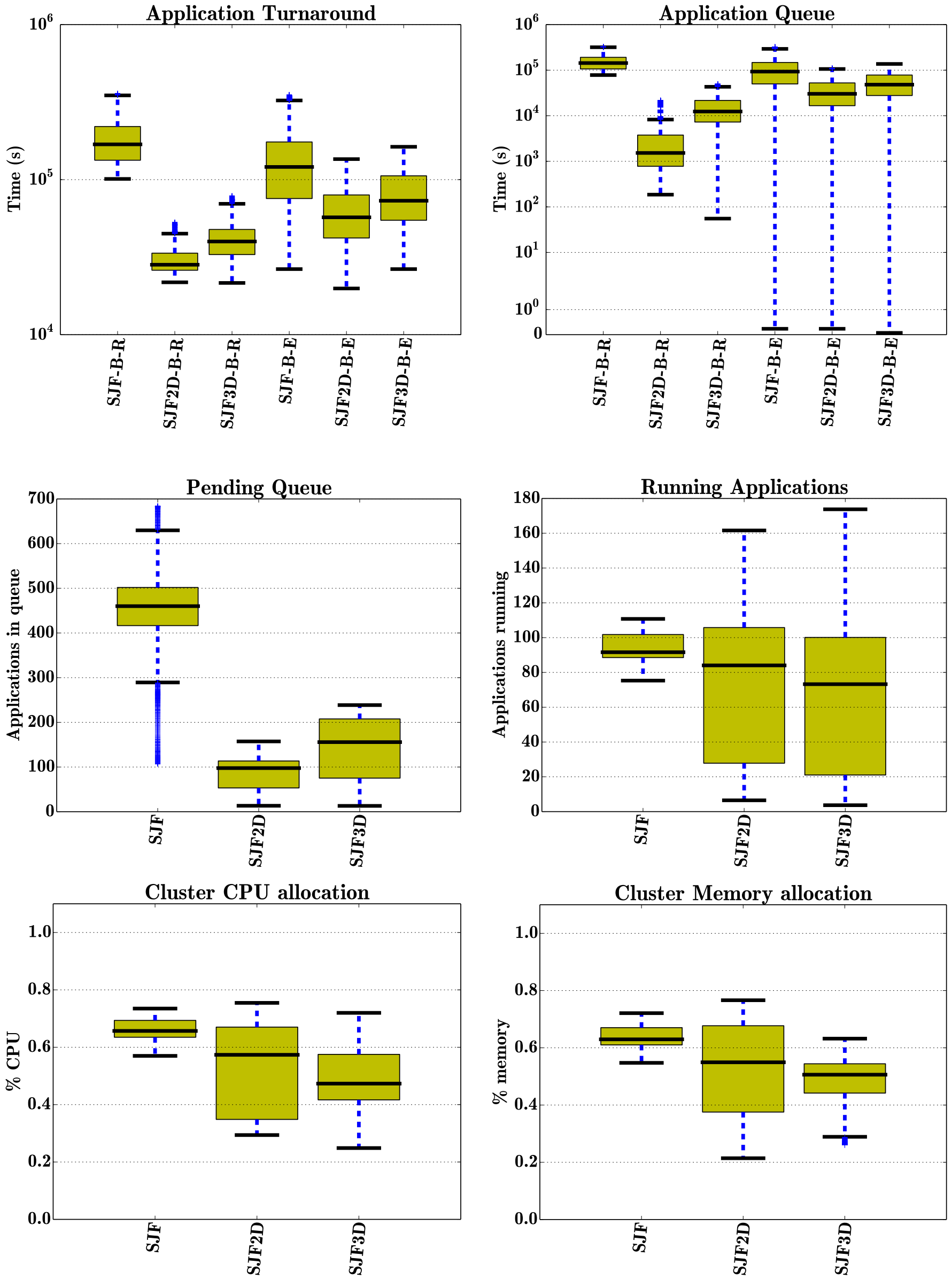

Figure 14: Comparison using different definition of size for SJF with a rigid scheduler. *B-E* stands for batch elastic and *B-R* stands for batch rigid applications.

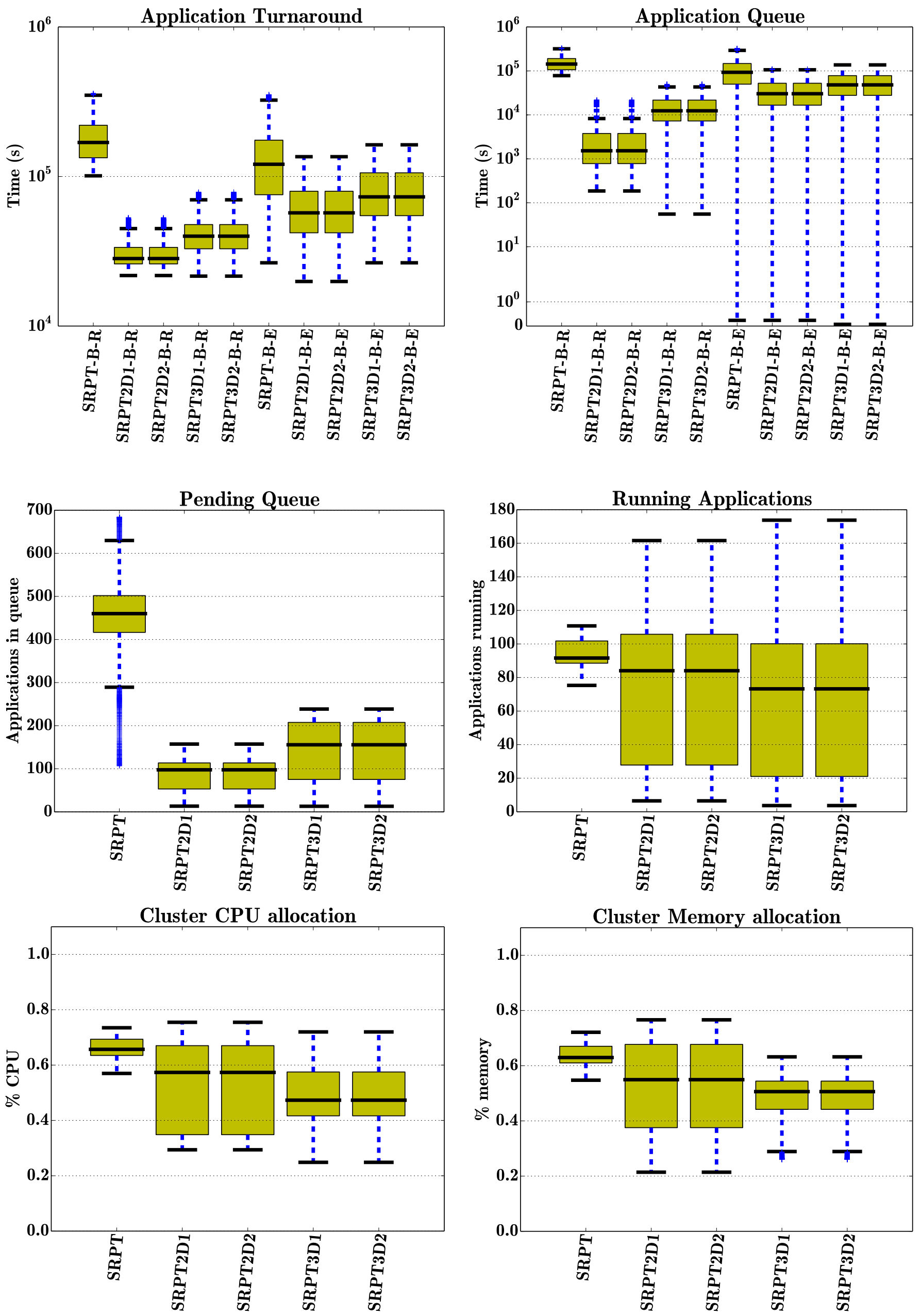

Figure 15: Comparison using different definition of size for SRPT with a rigid scheduler. *B-E* stands for batch elastic and *B-R* stands for batch rigid applications.

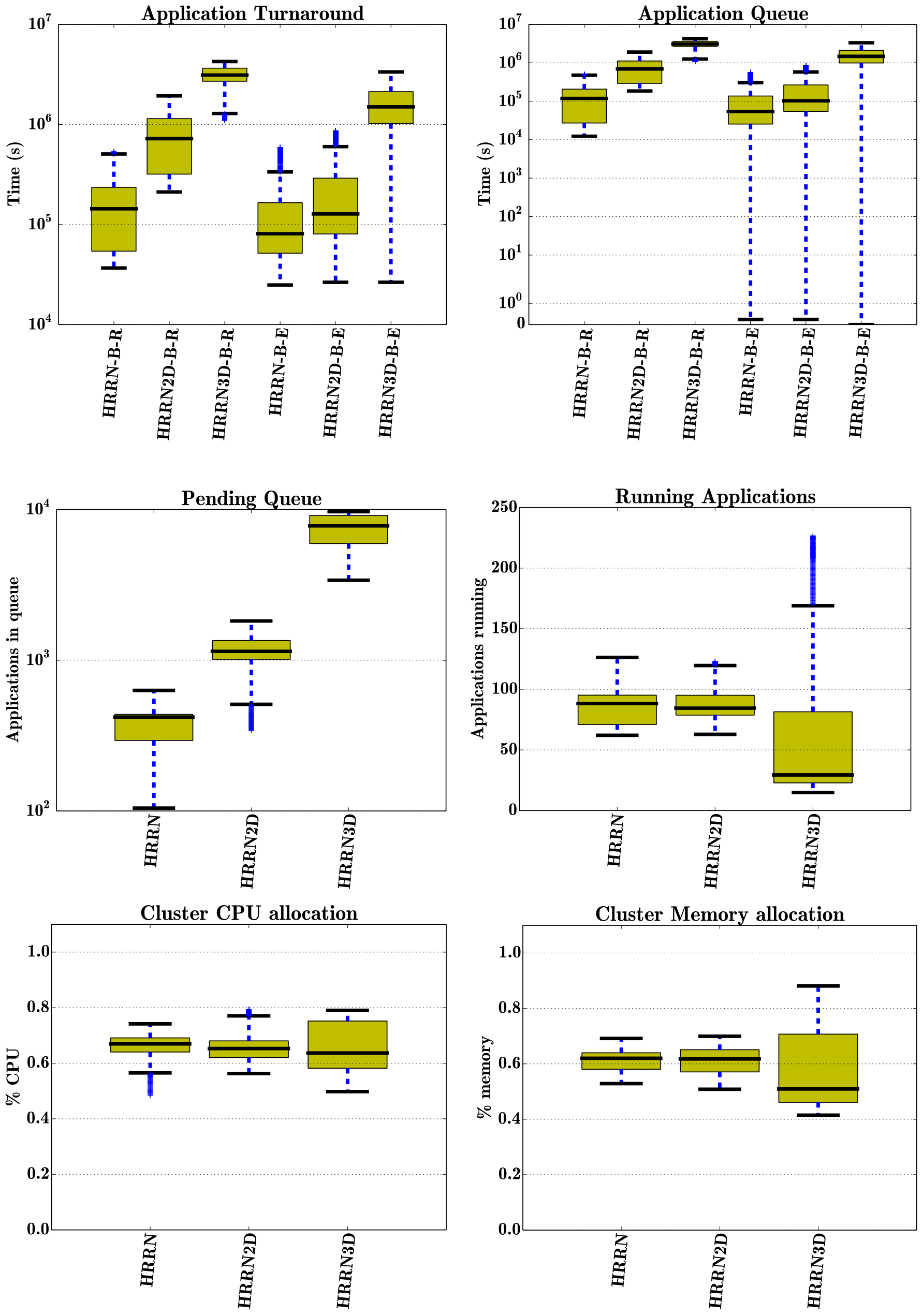

Figure 16: Comparison using different definition of size for HRRN with a rigid scheduler. *B-E* stands for batch elastic and *B-R* stands for batch rigid applications.

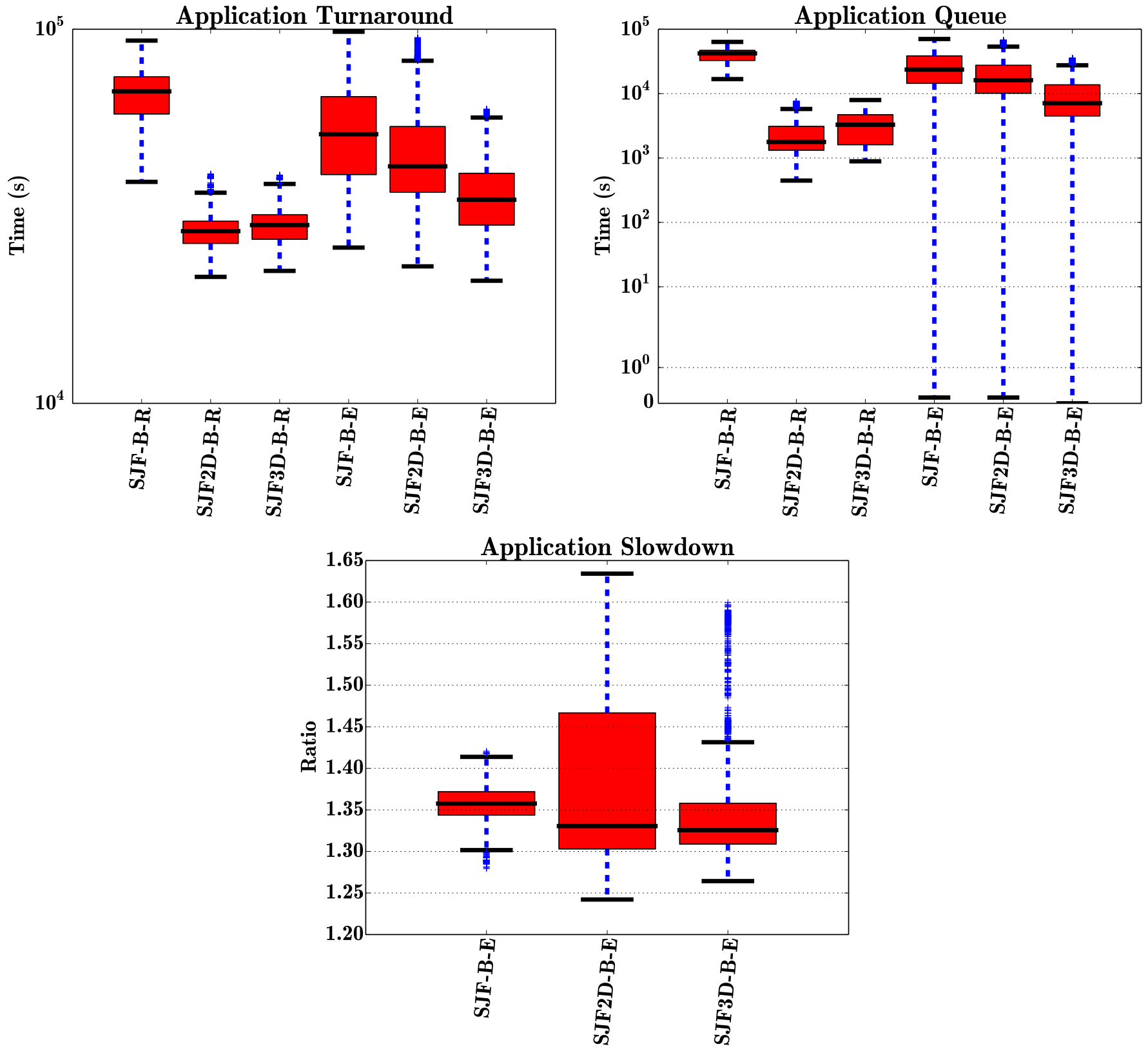

Figure 17: Comparison using different definition of size for SJF with a malleable scheduler. *B-E* stands for batch elastic and *B-R* stands for batch rigid applications.

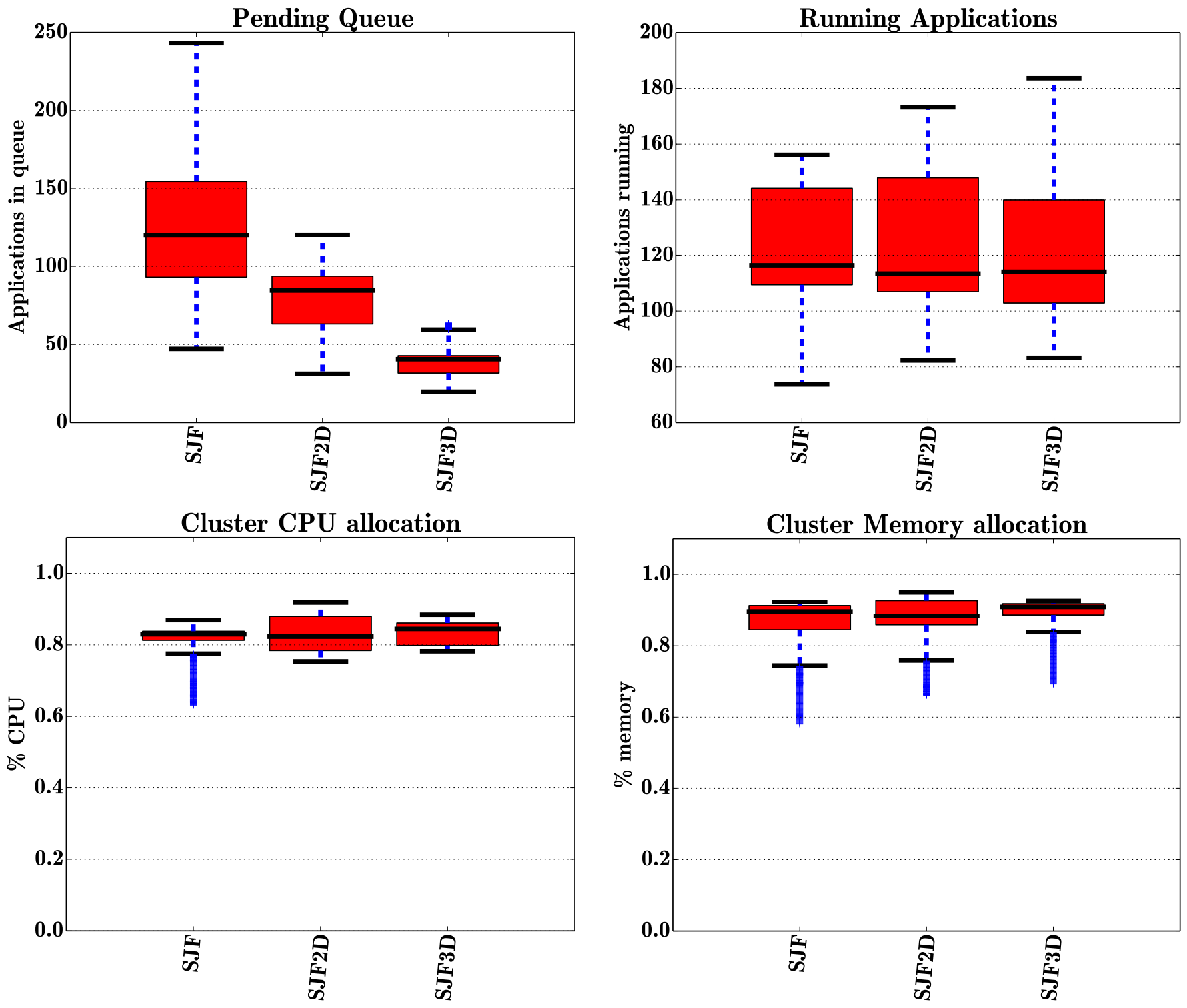

Figure 18: Comparison using different definition of size for SJF with a malleable scheduler. *B-E* stands for batch elastic and *B-R* stands for batch rigid applications.

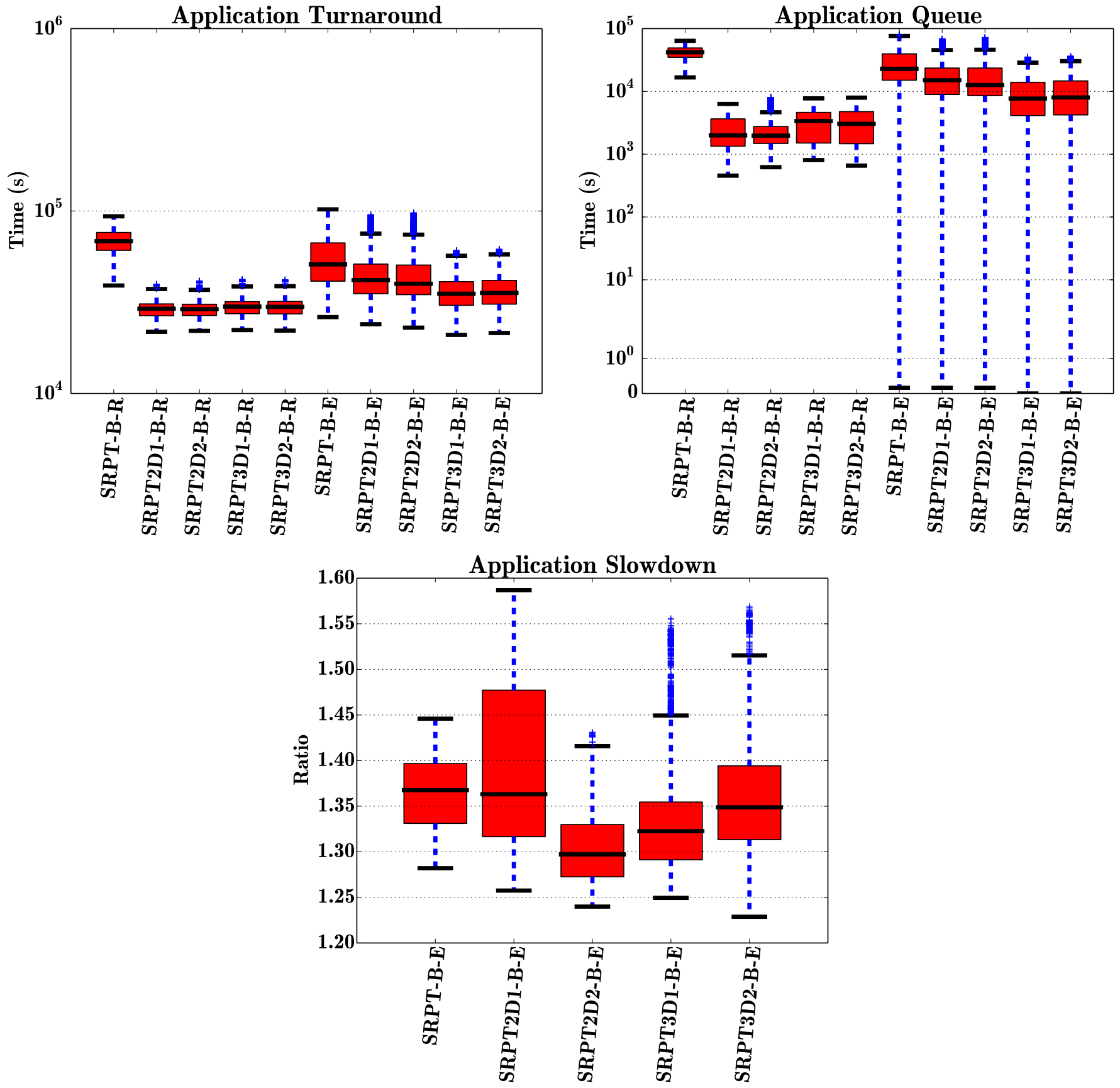

Figure 19: Comparison using different definition of size for SRPT with a malleable scheduler. *B-E* stands for batch elastic and *B-R* stands for batch rigid applications.

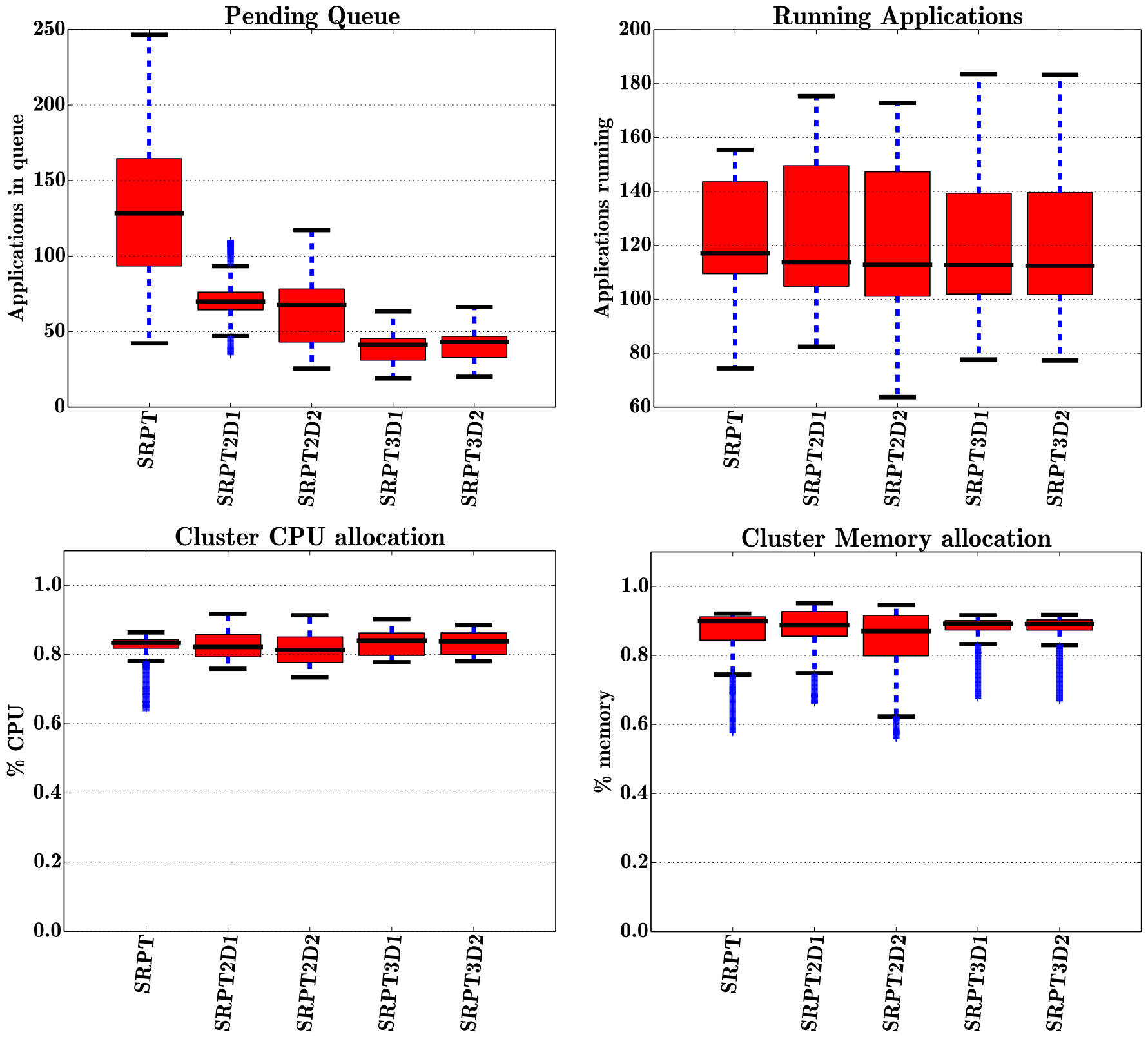

Figure 20: Comparison using different definition of size for SRPT with a malleable scheduler. *B-E* stands for batch elastic and *B-R* stands for batch rigid applications.

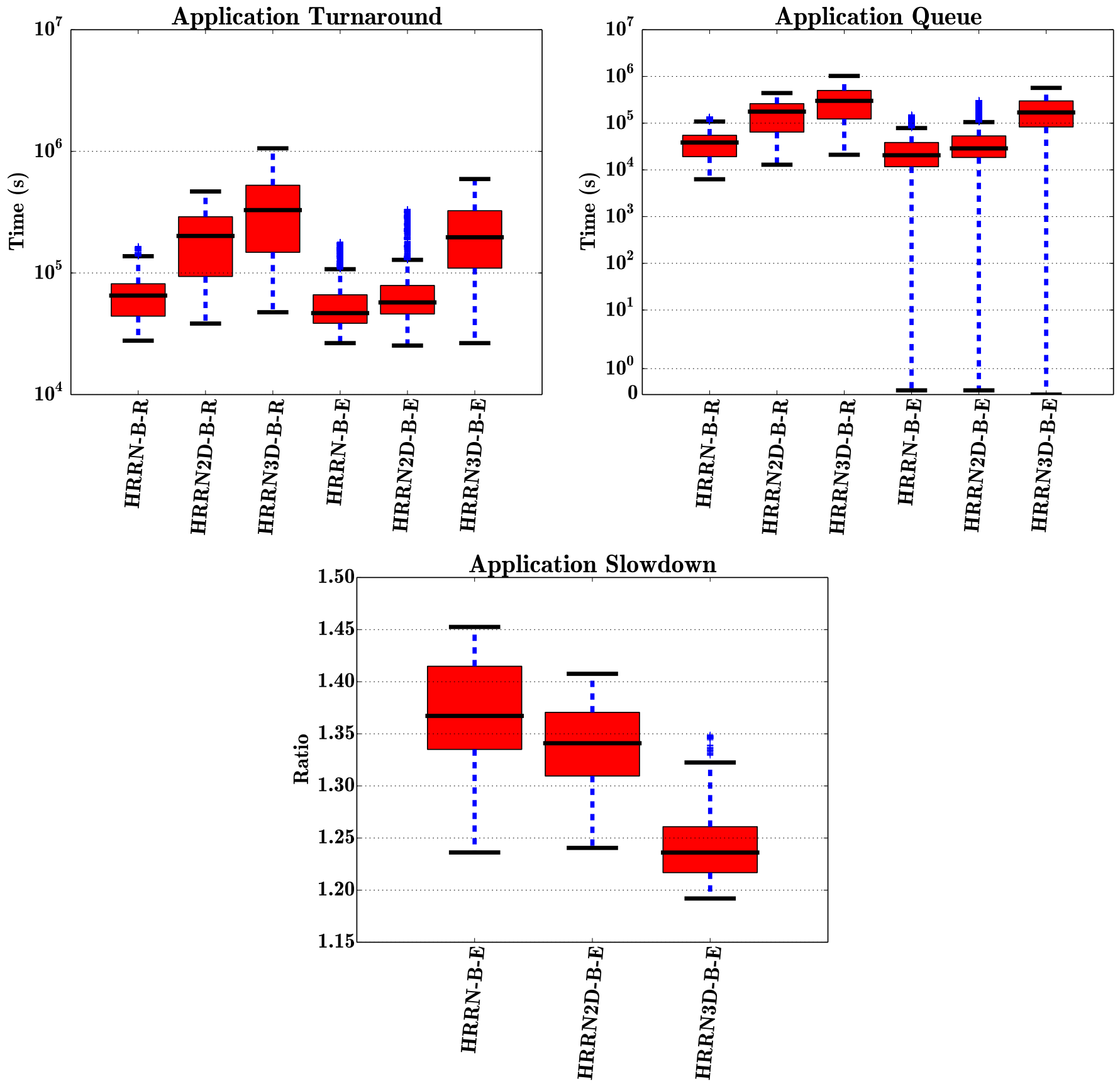

Figure 21: Comparison using different definition of size for HRRN with a malleable scheduler. *B-E* stands for batch elastic and *B-R* stands for batch rigid applications.

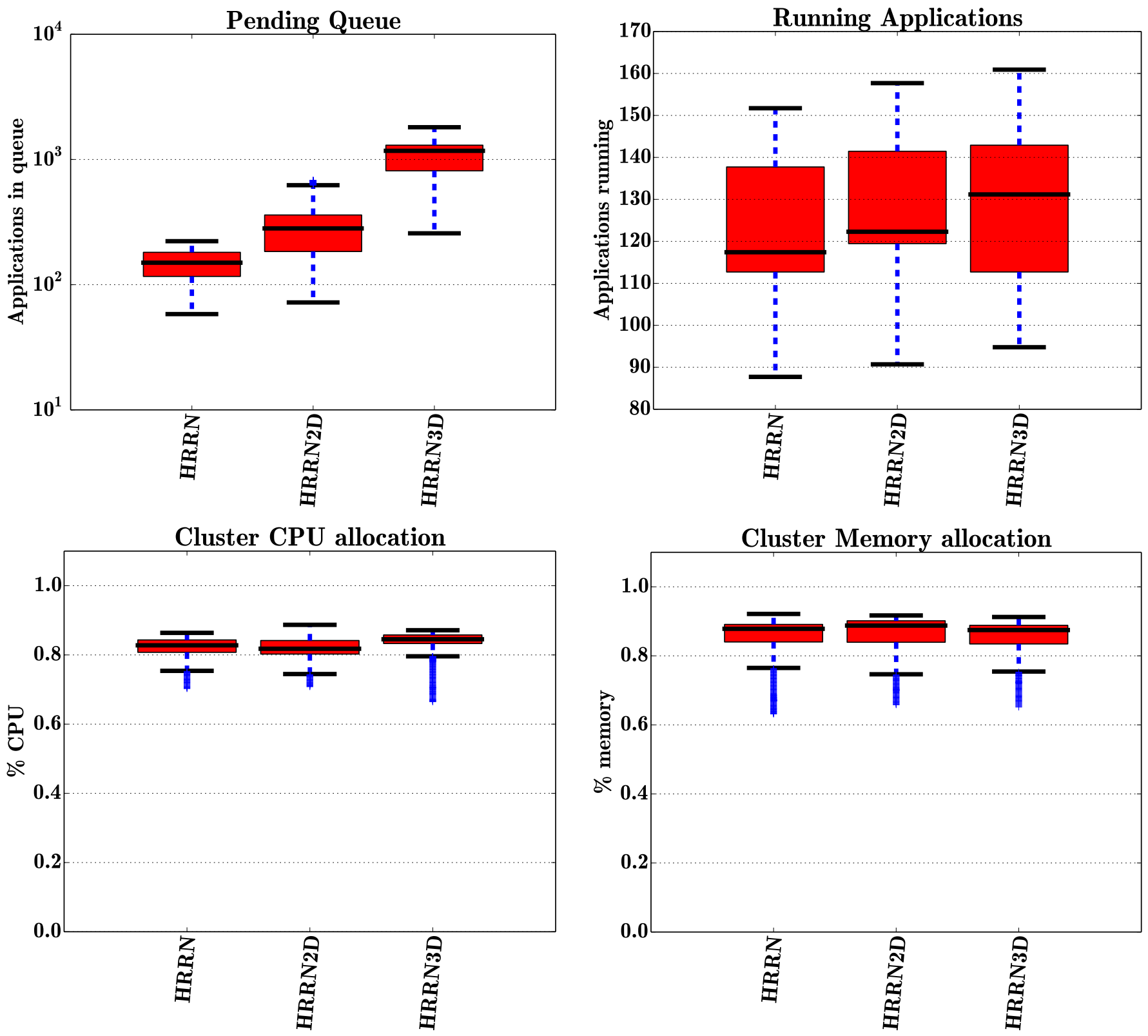

Figure 22: Comparison using different definition of size for HRRN with a malleable scheduler. *B-E* stands for batch elastic and *B-R* stands for batch rigid applications.

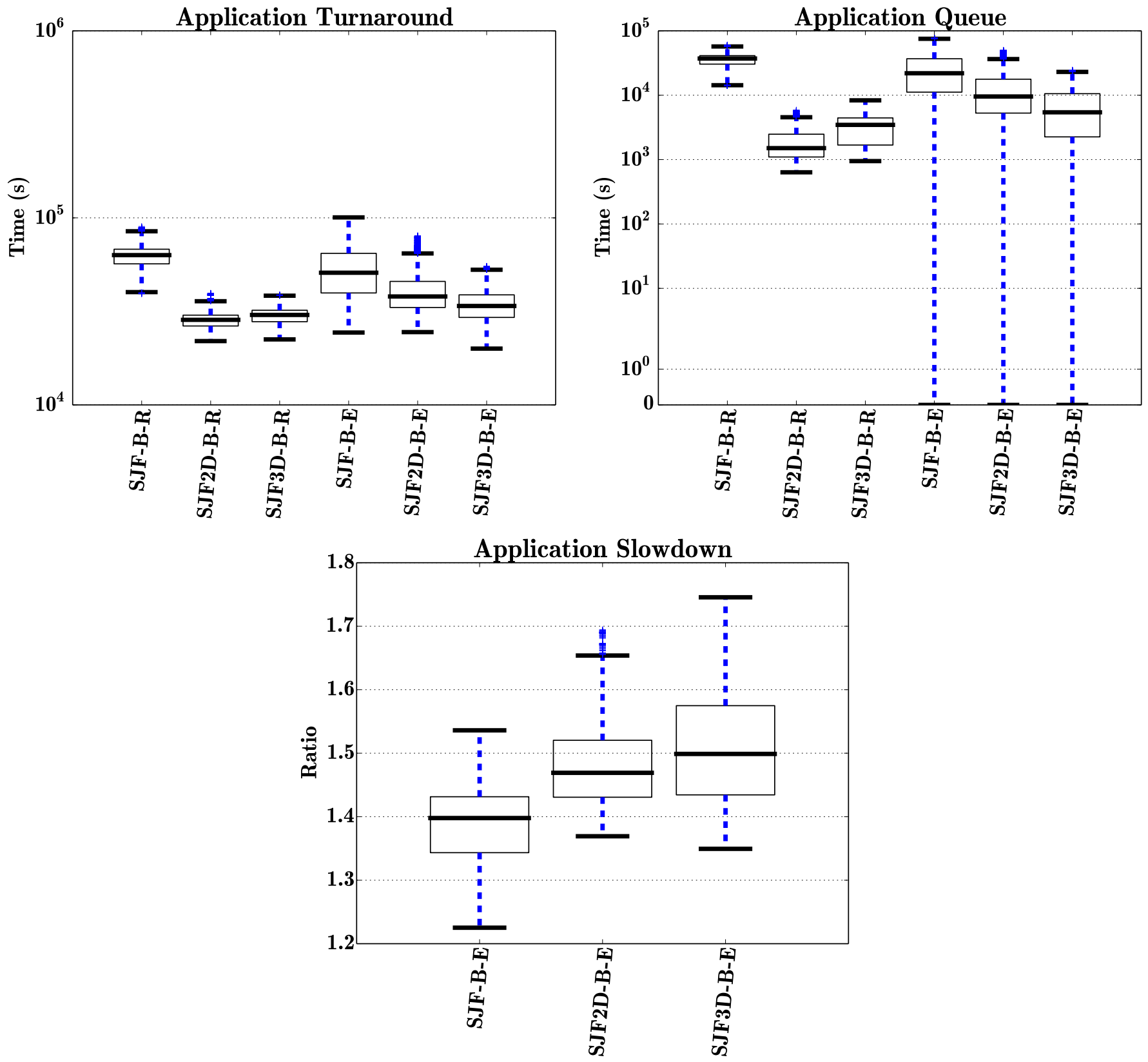

Figure 23: Comparison using different definition of size for SJF with our flexible scheduler. *B-E* stands for batch elastic and *B-R* stands for batch rigid applications.

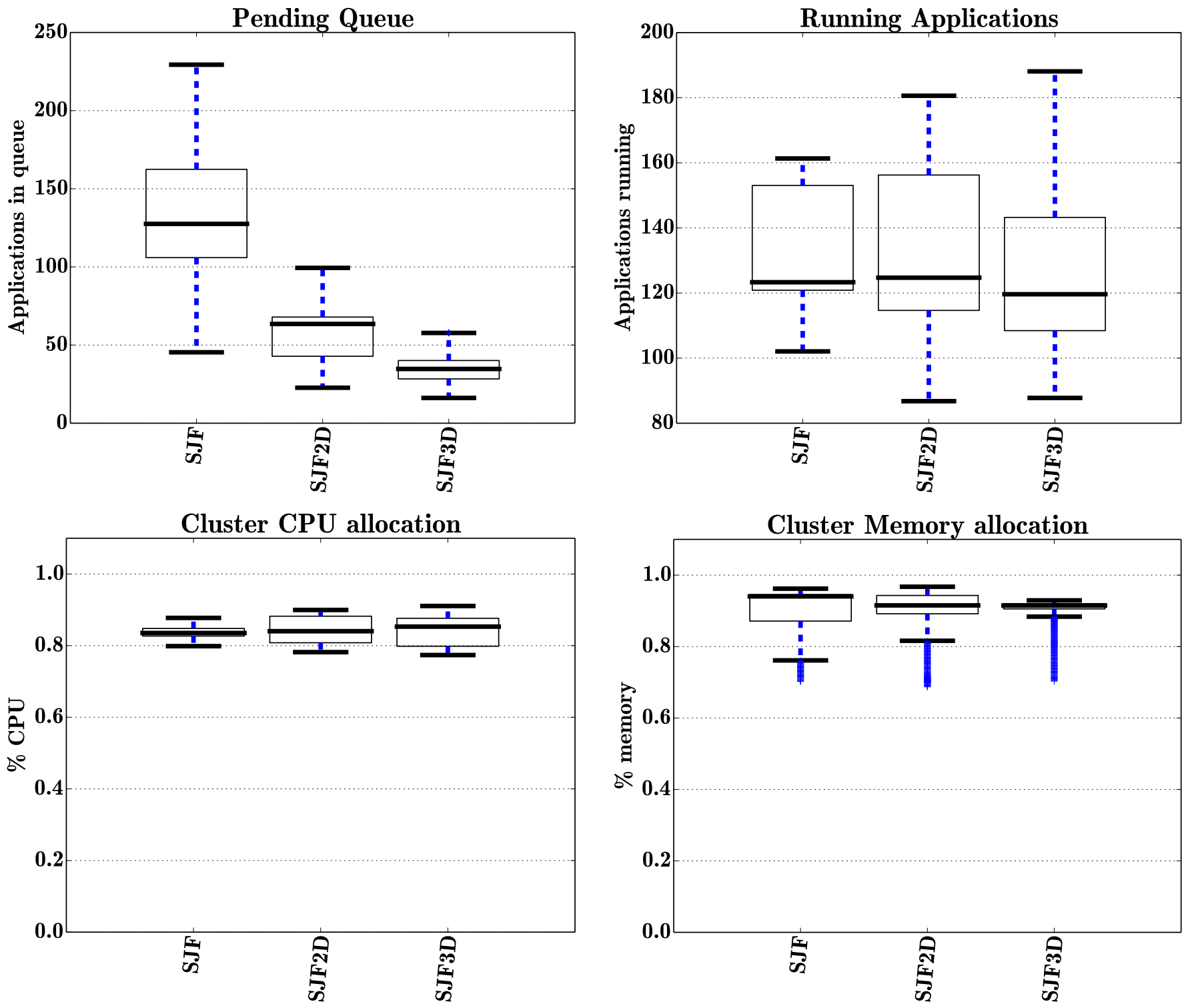

Figure 24: Comparison using different definition of size for SJF with our flexible scheduler. *B-E* stands for batch elastic and *B-R* stands for batch rigid applications.

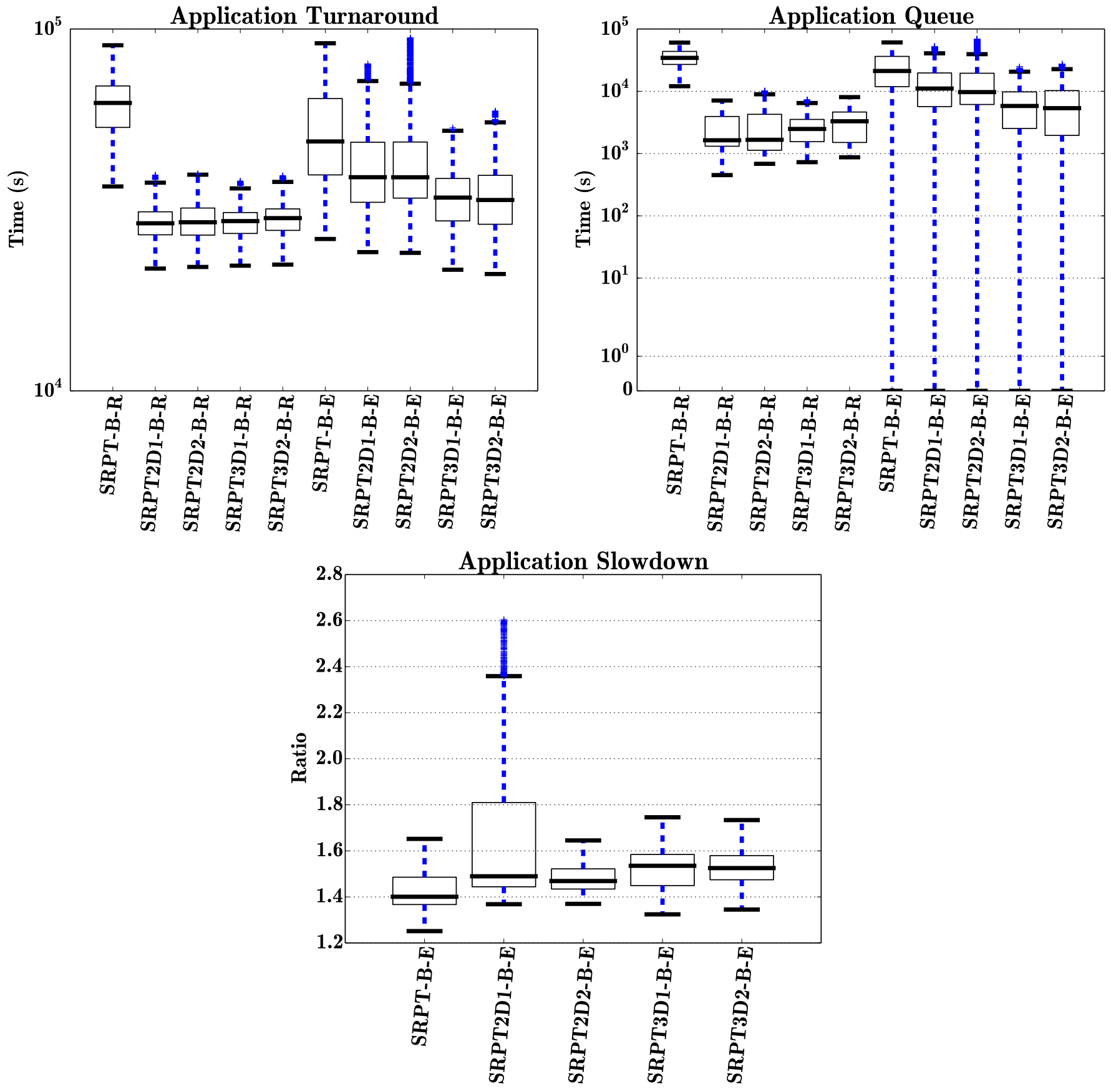

Figure 25: Comparison using different definition of size for SRPT with our flexible scheduler. *B-E* stands for batch elastic and *B-R* stands for batch rigid applications.

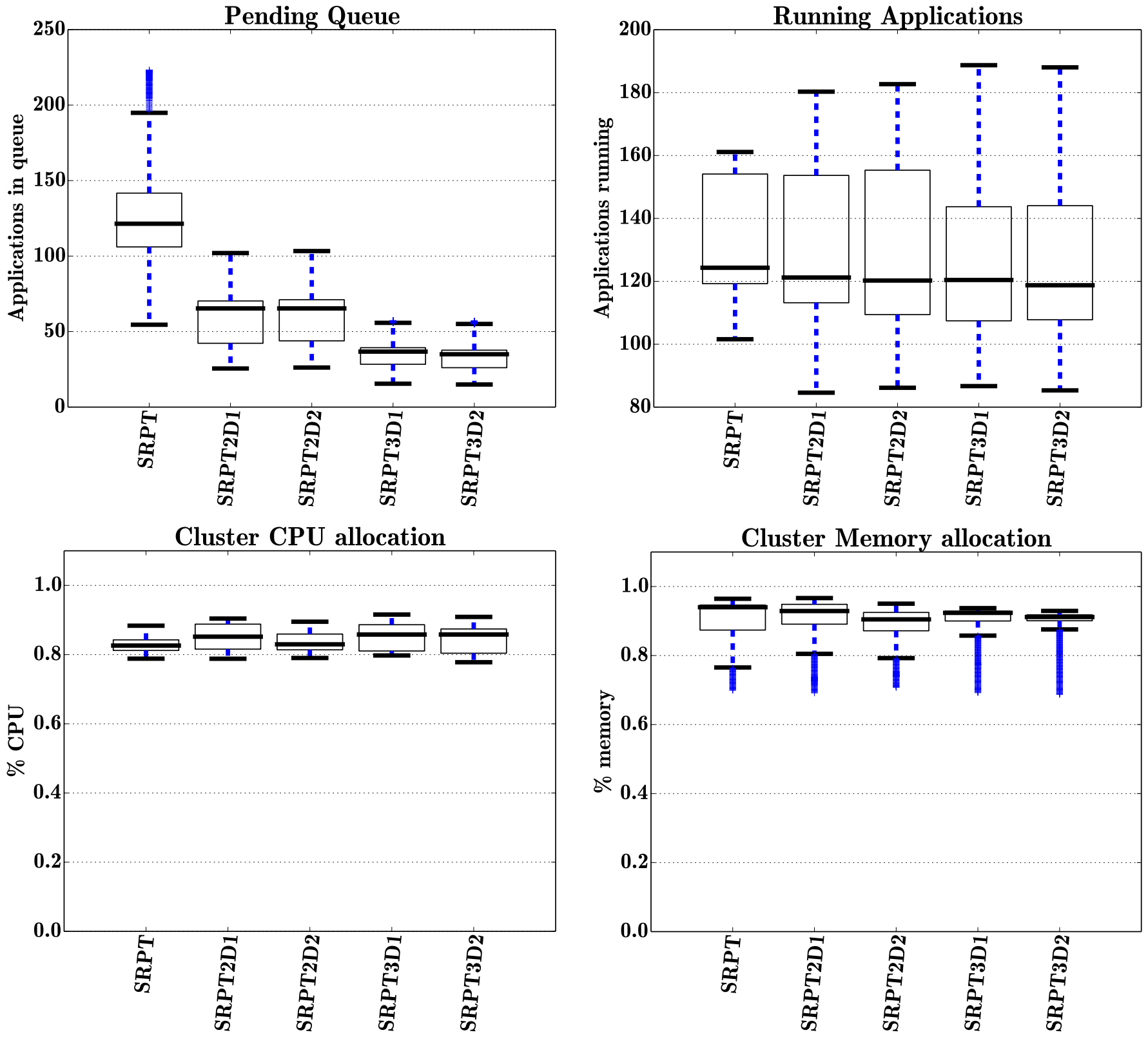

Figure 26: Comparison using different definition of size for SRPT with our flexible scheduler. *B-E* stands for batch elastic and *B-R* stands for batch rigid applications.

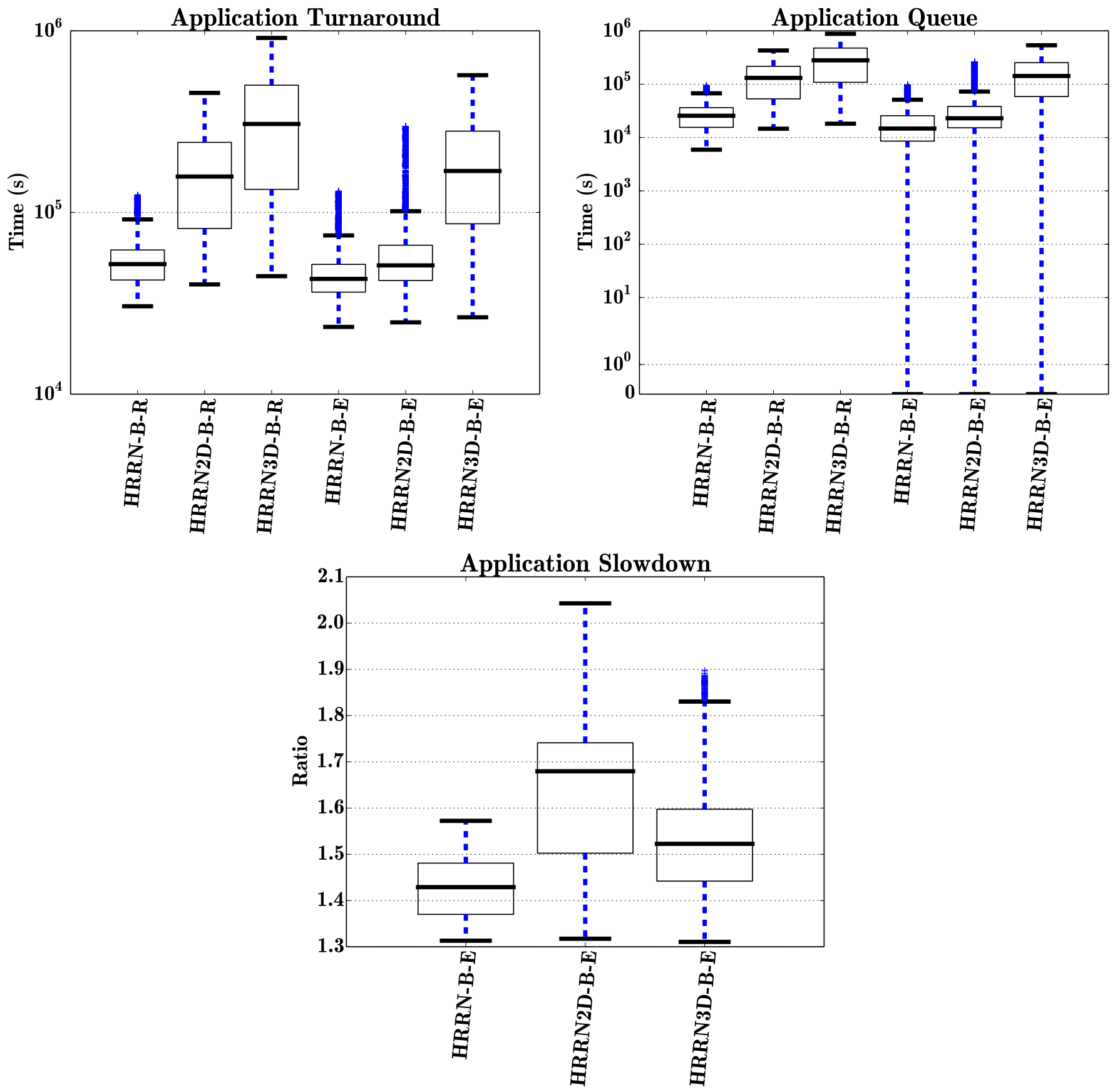

Figure 27: Comparison using different definition of size for HRRN with our flexible scheduler. *B-E* stands for batch elastic and *B-R* stands for batch rigid applications.

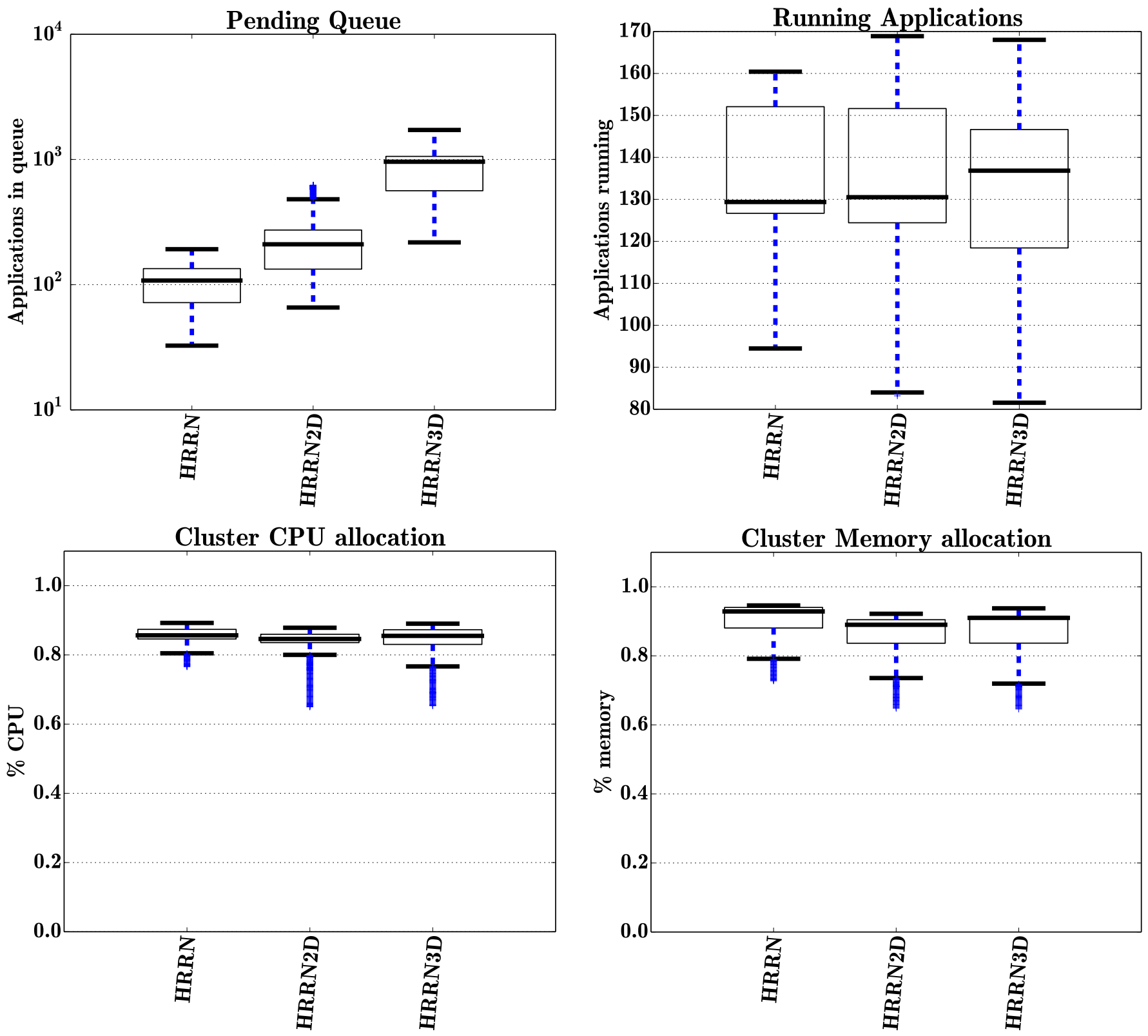

Figure 28: Comparison using different definition of size for HRRN with our flexible scheduler. *B-E* stands for batch elastic and *B-R* stands for batch rigid applications.

for rigid schedulers (Figs. 14 to 16); having a three dimensional size definition may improve the turn around of some policies compared to a unidimensional definition, but in all cases having a two dimensional definition is sufficient to obtain good results. The resource allocation, however, decreases a bit in a rigid system as more information is added; it will allows big-but-short (big resource requirement but short lived) applications to block the queue thus, increase queue times and reduce resources utilization in the cluster. Since our flexible scheduler allows only the indispensable (inelastic) services to be launched, it can overcome the problems that a rigid has with a three dimensional definition and, as a result, we can see that it performs better than a one or two dimensional definitions.

*In conclusion, adding more information to the size does not always improve the turn around time in a rigid scheduler because some big-but-short applications can block the queue. However, with a malleable and with our flexible scheduler this can be avoided and we can have lower turn around time and higher cluster resources utilization with a three dimensional definition of size. We would like to stress more the fact that the malleable scheduler is something widely adopted in theory, but not in use real systems.*

## 4.4 Impact of different workloads

In Sections 4.2 and 4.3 we show the benefits on turn around time of a flexible scheduler solutions – that split the services of an application in core and elastic – compared with a *rigid* scheduler that does not. It is intuitive to think that the results highly depend on the workload; on the one hand, more percentage of elastic applications lead to higher benefit for our flexible scheduler; on the other hand, more inelastic (composed by only core components) applications will move the results much closed to the rigid. However, in the worst case our flexible scheduler will perform exactly the same as the rigid; the flexible scheduler will have to allocate the requested resources at one, therefore its behavior is the same as the rigid. Table 3 shows the results of a simulation with a completely inelastic workload; the results show that the average turnaround is exactly the same.

*In conclusion, our flexible scheduler does not introduce any overhead and, in the worst case, will not perform worst than a rigid.*

| | FIFO | PSJF | SRPT | HRRN |
|---|---|---|---|---|
| Rigid | 1,317,315.92 | 135,964.50 | 135,964.50 | 407,952.05 |
| Hybrid | 1,317,315.92 | 135,964.50 | 135,964.50 | 407,952.05 |

Table 3: Turn around time (in second) between a rigid and our flexible scheduler with a workload composed by only inelastic applications.

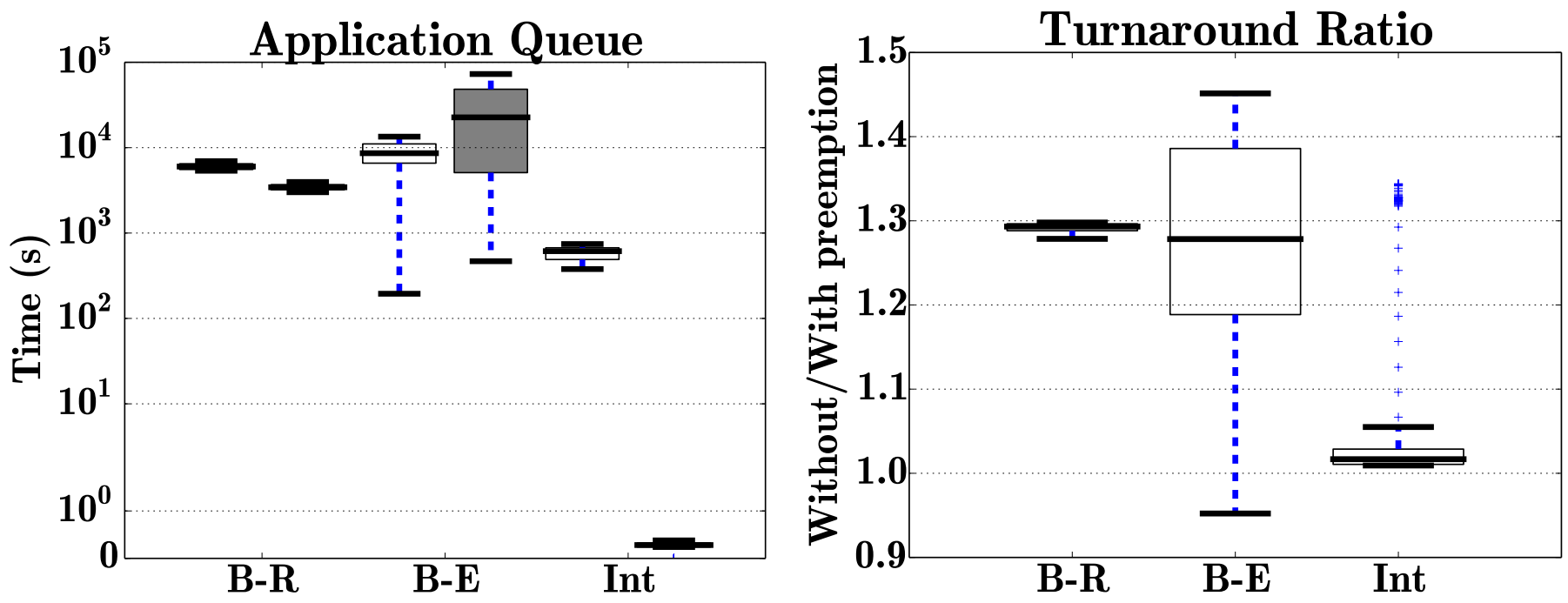

Figure 29: On the left, comparison of queuing time distributions between scheduling with and without preemption. White boxes (left box of every pair) correspond to a *non-preemptive* system, gray boxes to our preemptive algorithm. On the right, turnaround ratio distributions between scheduling with and without preemption. *B-E* stands for batch elastic applications, *B-R* stands for batch rigid applications and *Int* is for interactive applications.

### 4.5 Preemption

We turn now our attention to the full workload we defined in Section 4.1, including *interactive* applications. Preemption is used when a high-priority, interactive application requires resources: this applies both to manually set priorities (e.g., in a FIFO policy) and to size-based priorities. In particular, we report results for the preemptive version of the SRPT policy.

Figure 29 shows the most relevant percentiles of the distribution of application queuing times, grouped by application type (both cases of batch and interactive applications), with and without our preemption mechanism. Globally, preemption does not subvert the perceived system responsiveness. However, interactive applications under preemptive scheduling enjoy roughly two orders of magnitude less queuing times. Users do not wait for few dozens minutes but only few seconds, for their interactive application to start. As a consequence, elastic batch applications pay with more variability (but stable for the median case) in queuing times.

Since our simulator does not account for real work done by applications, the preemption mechanism does not have any effect on the work that has been done by preempted components. In practice, our current preemption mechanism would instead suppress work done by elastic services, if preempted. Studying new preemption primitives, e.g. by suspending Linux containers, is part of our research agenda: this is the main reason why our current prototype implementation lacks support for preemption.

Finally with Figs. 30 to 32 we show the results for all the policies with their different definition of size that we identified in Section 4.3; globally, the considerations made before for the SRPT policy can be applied for the other policies as well. However, we can see a drop in the resource utilization and an increase in the size of the pending queue when using a preemptive system; this is because we are no longer respecting the order imposed by the policy, but we are prioritizing interactive applications that have a much longer duration compared to the batch, thus a lot of shorter applications are forced to wait.

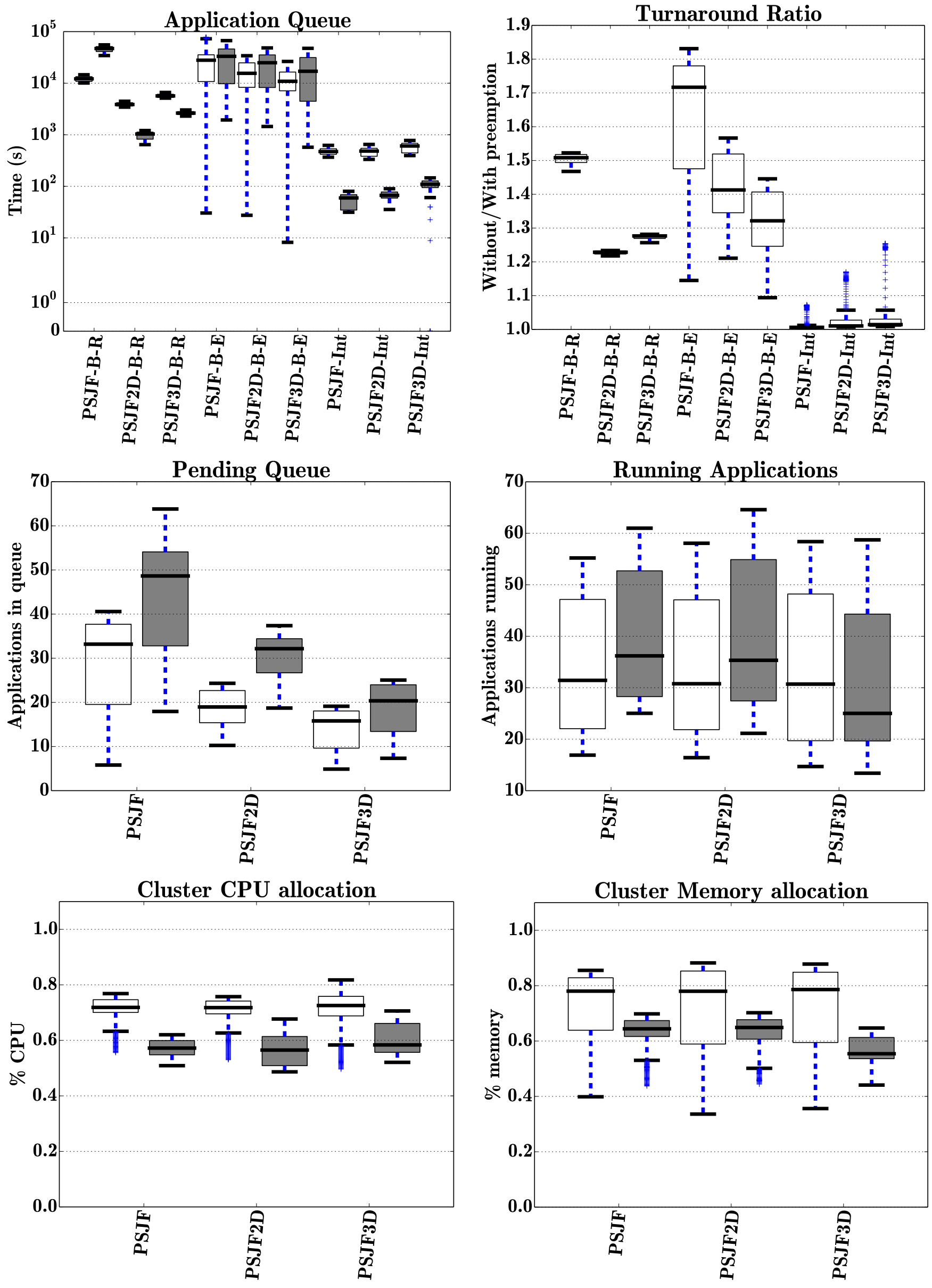

Figure 30: Comparison between a system with and without preemption for PSJF and its different definition of size. White boxes (left box of every pair) correspond to a *non-preemptive* system, gray boxes to our preemptive algorithm. *B-E* stands for batch elastic applications, *B-R* stands for batch rigid applications and *Int* is for interactive applications.

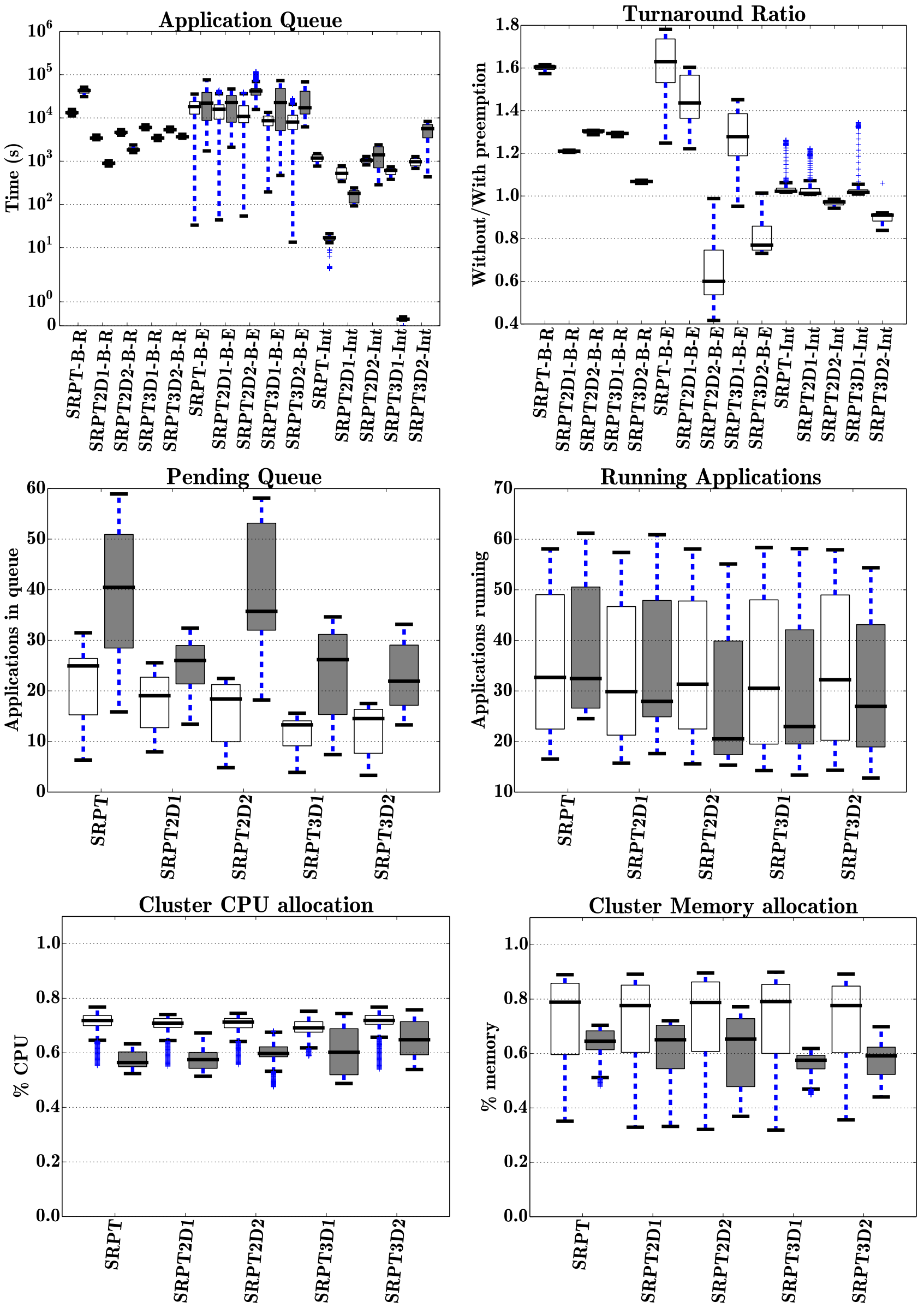

Figure 31: Comparison between a system with and without preemption for SRPT and its different definition of size. White boxes (left box of every pair) correspond to a *non-preemptive* system, gray boxes to our preemptive algorithm. *B-E* stands for batch elastic applications, *B-R* stands for batch rigid applications and *Int* is for interactive applications.

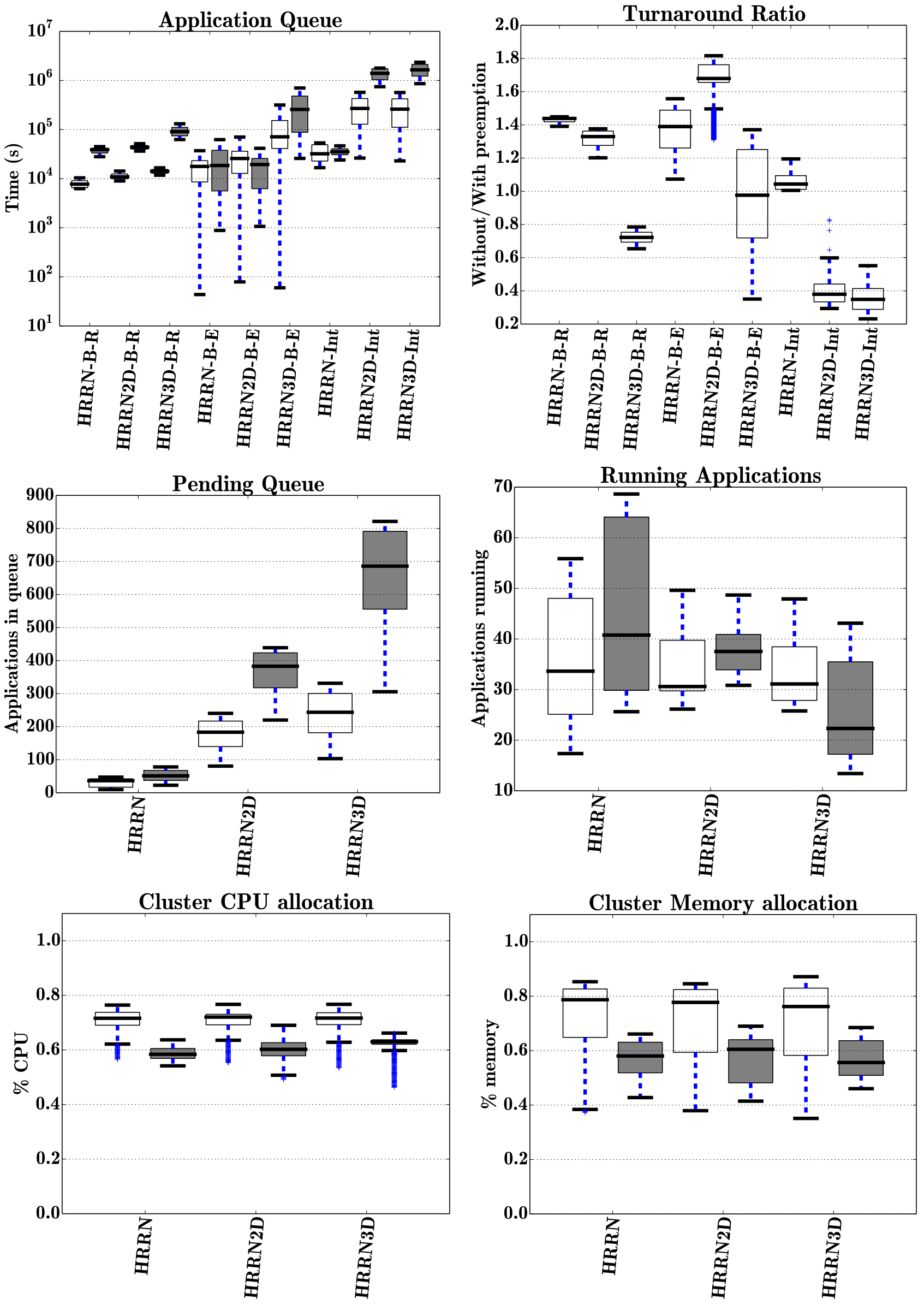

Figure 32: Comparison between a system with and without preemption for HRRN and its different definition of size. White boxes (left box of every pair) correspond to a *non-preemptive* system, gray boxes to our preemptive algorithm. *B-E* stands for batch elastic applications, *B-R* stands for batch rigid applications and *Int* is for interactive applications.

# 5 Implementation: The Zoe system

Next, we describe Zoe[5], the system we have built to materialize the concepts developed earlier.

Zoe allows defining *analytics* applications and schedule them in a cluster of machines. It is designed to run on top of an existing low-level cluster management system, which is used as a back-end to provision resources to applications. Raising the level of abstraction to manipulate analytics application is beneficial for users and ultimately to the system design itself: application scheduling decisions can be taken with a small amount of state information, and do not happen at the same (extremely fast) pace as low-level task scheduling. Next we overview Zoe's design, and provide relevant details for the subject of this work.

**Zoe applications.** In Zoe, the concepts introduced in Section 2 take the form of simple JSON description files that follow a high-level configuration language (CL) to specify applications, frameworks and components with their classes (core or elastic), resource reservations and constraints. The CL is simple and extensible: it aims at conciseness and, with framework templates, can be used by "casual" and "power" users [10].

The key aspect that determines the application type (batch, interactive, or any new type) is the way application life-cycle is managed. This is determined by a flexible attribute, reminiscent of a "command line", which allows passing runtime configuration options, user-defined arguments and environment variables, as well as setup and cleanup procedures. For application design, the "command line" attribute requires minimal knowledge of the frameworks that constitute an application. As an example of the simplicity and effectiveness of the Zoe CL, building a batch application for the distributed version of TensorFlow [27] only requires tens of lines of CL. The most important attribute is the "command line", which is required to run a TensorFlow program, *i.e.,* `python $TF_PROGRAM $PS_HOSTS $WK_HOSTS program-args`. Environment variables are appropriately handled by Zoe, including information unknown at scheduling time (e.g., host names).

A note on application failures is required. Any failure of an elastic component is practically harmless, whereas core component failures imply application failure. An area of future work is to exploit failure tolerance mechanisms available from some back-ends (e.g., Kubernetes) to steer application-level failure

[5]Zoe, `https://zoe-analytics.eu/`, was conceived in August 2015, named after the biggest container boat in the world, which, has touched sea [37] in the same period of time. In this work, we omit several implementation details that stem from our continuous effort to extend Zoe.

tolerance modes.

**Zoe back-ends.** The main design idea of our system is to hide the complexities of low-level resource provisioning from application scheduling and exploit an existing cluster management system, for which many alternatives exists. Currently, Zoe builds on top of Docker Swarm [23], and uses it to achieve a series of objectives we list below:

- *Orchestration*: Zoe interacts with all the machines in a cluster using the Docker orchestration API (known as Swarm [23]), which governs the behavior of the Docker engine [38] deployed in each machine. Thus, Zoe manages to distribute the necessary binaries for the components of an application that is scheduled for execution, their configuration, life-cycle, and provisioning.
- *Dependency management*: Zoe applications materialize as a series of Docker images, which contain all dependencies and external libraries required for an application to run. Zoe applications can be built from *community-provided* or custom Docker images of existing frameworks.
- *Resource isolation*: framework components specified in an application run in Linux containers, which are managed by a Docker engine. We also use the Docker engine to achieve memory allocation, whereas CPU partitioning is left to the machine OS. This means, we have a one dimensional packing problem.
- *Resource matching*: application descriptions include resource constraints. When an application is scheduled for execution, Zoe instructs the back-end to adhere to component constraints when provisioning the relevant Docker images with framework binaries, as determined by the virtual assignment obtained by Algorithm 1.
- *Naming and networking*: the services for application components to cooperate in producing useful work, and to interact with the outside world are an important aspect to consider when choosing an appropriate back-end for Zoe. We use Docker networking, but we also have developed our own service discovery mechanism to allow a more flexible application configuration and deployment.

**Zoe architecture.** Although Zoe is separated in several modules, it does not require any cluster-wide installation, because it uses its back-end to interact with the cluster.

The Zoe *master* polls a high-fidelity view of the cluster state through its back-end, whenever the scheduler is triggered, and stores it into a *state store*, backed by a PostgreSQL database. The state store also holds applications state, which is modeled as a simple state-machine. Because Zoe handles high-level objects (applications), the strain on the system is minimal: the rate of scheduling decisions scales well even with heavy workloads. The virtual assignment procedure avoids *application interference* by construction because it considers requests in sequence, according to their priority. The virtual assignment is imposed on the back-end, using its API.

The Zoe *client API* handles REST calls that mutate the system state, or that can be used to monitor the system behavior. Command-line and web interfaces allow users and administrators to interact with the system and the cluster.

The Zoe *scheduler* implements the algorithm described in Section 3. When an application is submitted, the Zoe master creates an entry in the application state store, and adds it to a *pending queue*. Our system allows plugging several *scheduling policies* to manage the pending queue, ranging from simple to sophisticated size-based policies. Such policies determine which application is granted "access" to cluster resources: to this end, the scheduler uses the cluster state store to simulate possible deployments before accepting an application. Framework components underlying an application are scheduled according to their type. The scheduler strives at making sure the application selected for execution can make progress as soon as resources are allocated to it. The Zoe *monitoring* module uses the Docker event stream to update the state of each application component running in the system.

Currently, the Zoe system supports a naive *preemption mechanism*: entire applications can be killed upon a command. The finer strategy described in Section 3 and Section 4 is currently under development.

Finally, although Zoe supports many data sources and sinks, we report experiments using a HDFS cluster to store input data to applications, and CEPH volumes to store application-specific logs.

# 6 Experiments with Zoe

Our goal now is to perform a comparative analysis of two generations of Zoe: the first, implementing a rigid scheduler, as for the baseline, the second with the flexible scheduler we present here. In our experiments, we replay the exact same workload trace for both generations. Each trace takes about 3 hours from the first submission to the last. During our experiments, no other user was allowed to submit jobs to Zoe.

**Workload.** We use two representative *batch application templates*, including: 1) an elastic application using the Spark framework; 2) a rigid application using the TensorFlow framework. Following the statistical distribution of our historical traces, we set our workload to include 80% of elastic and 20% of rigid applications, for a total of 100 applications. Application inter-arrival times follow a Gaussian distribution with parameters $\mu = 60$ sec, and $\sigma = 40$ sec, which is compatible with our historical data. More specifically, using the elastic application templates, we run two use cases. First, an application to induce a random-forest regression model to predict flight delays, using publicly available data from the US DoT.[6] Second, a music recommender system based on alternating least squares algorithm, using publicly available data from Last.fm[7].

[6]http://stat-computing.org/dataexpo/2009/the-data.html

[7]http://www-etud.iro.umontreal.ca/~bergstrj/audioscrobbler_data.html

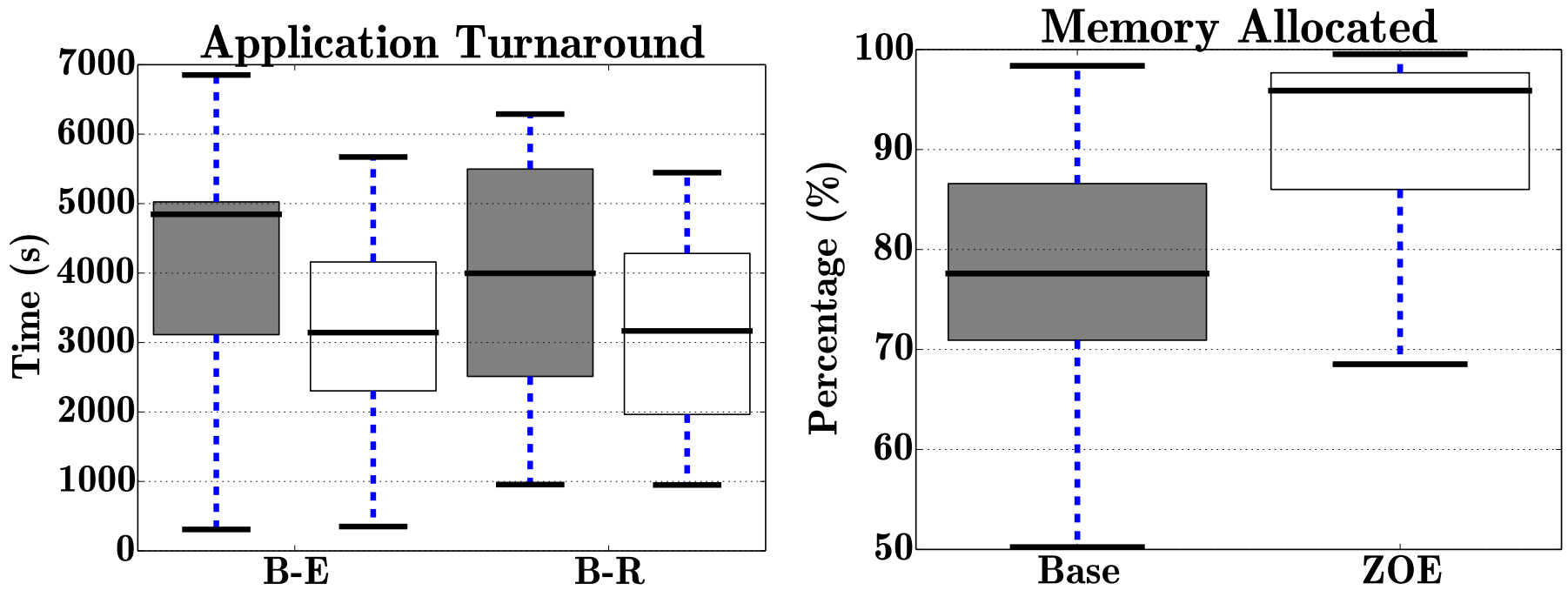

Figure 33: Comparison of turnaround time distributions using the FIFO discipline. White boxes (right box of every pair) correspond to the second generation of Zoe that implements our algorithm. *B-E* stands for batch elastic and *B-R* stands for batch rigid applications.

Both applications have two different requirements in term of memory for each elastic component. The random-forest regression model has 3 core components and then 32 elastic components of 16GB or 8GB RAM each; every elastic component uses 1 CPU. The music recommender system has 3 core components and then 24 elastic components of 16GB RAM or 8GB each; every elastic component uses 6 CPU. Instead, using the rigid application template, we train a deep Gaussian Process model [39], and use both a single-node and a distributed TensorFlow program, requiring 1 and 10 workers (and 5 parameter servers) each with 16GB of RAM.

**Experimental setup.** We run our experiment on a platform with ten servers, each with two 16-core Intel E5-2630 CPU running at 2.40GHz (total of 32 cores with hyper-threading enabled), 128GB of memory, 1Gbps Ethernet network fabric and ten 1TB hard drives. No GPU-enabled machines are available in our platform, at the moment. The servers use Ubuntu 14.04, Docker 1.11 and the standalone Swarm manager. Docker images for the applications are preloaded on each machine to prevent container startup delays and network congestion.

**Summary of results.** Using the FIFO scheduling policy, we compare the two generations of Zoe according to the distributions of application turnaround times, as shown in Figure 33 (left). The behavior of the two systems indicate a clear advantage for our approach: the median turnaround times are 37% and 22% lower, for elastic and rigid applications respectively. Note also that the tails of the distributions are in favor of our approach.

Overall, the new generation of Zoe that implements the flexible scheduler is more efficient, with a 20% improvement, in allocating and packing applications, as illustrated in Figure 33 (right), where we show the ratio of the distribution of allocated over available resources.

Finally, we present results concerning a low-level metric that measure the application *ramp-up time*, *i.e.,* the time it takes for applications scheduled for running, to receive their allocations and start producing work. Zoe achieves a container startup time, including placement decisions, of $0.90 \pm 0.25ms$. Full-fledged applications, made by several containers, only take few seconds to start, which is a compelling property, especially when compared to existing solutions such as Amazon EMR.

# 7 Related Work

While we cannot do justice to the richness of the scheduling literature, in this section we organize related work in three groups. The "competitors" are existing, mature systems that could be considered sufficient to address our problem statement, at a first sight. The second group includes recent works in the systems research literature, while the third covers works on scheduling at the task level.

Many "competitor" systems have been designed to cope with the problem of sharing cluster resources across a heterogeneous set of applications, some of which can be tweaked to achieve the goals we set in this work. For example, Yarn [21] and Mesos [8] have been among the first to enable multiple frameworks to coexist in the same cluster: usage of these "two-level" schedulers yield a big improvement as compared with monolithic approaches to resource scheduling. Originally designed for analytic frameworks, such systems deal with the scheduling of low-level processing tasks. Recently, more general approaches address the problem of cluster-wide resource management: Omega [9], then Borg [10] (and Kubernetes [22]) reason at the "container" level, and are optimized to achieve efficient placement and utilization of cluster resources, when absorbing a very heterogeneous workload. This latter includes a majority of *long-running services*, which power Web-scale, latency-sensitive applications. Additionally, container orchestration frameworks, such as Docker Swarm [23], also provide efficient and scalable solutions to the problem of scheduling (that is, placing and provisioning) containers in a cluster. Our work relies on many of the above systems, and can use them as a back-end to support scheduling of high-level applications rather than provisioning low-level containers. Existing auxiliary deployment tools such as Aurora [40] and Docker Compose [41], do not address scheduling problems.

In the systems research literature, we find several inspiring system designs, with ideas that can be "borrowed" to further extend our system prototype and research scope. For example, Koala-f [14] tackles dynamic resource allocation problems, which manifest in our case with "idle" interactive applications. Similarly to HCloud [13], we believe important to break the reservation paradigm, and allow users to express performance requirements rather than machine-level resource counts. This can be easily included in an application description, but requires developing additional components to infer statistical models of application properties based on systems observations, as done for example in Tarcil [19], Paragon [15] and Quasar [16]. Overall, by focusing on a higher-level of abstraction, the focus of our work is to address a rather abstract scheduling problem:

our implementation indicates that our ideas work in practice and also bring tangible benefits.

Finally, many works address the problem of low-level task scheduling. Such schedulers are designed to support a specific "data-flow" programming model, but many of their design choices can also be used at a higher level. For example, Tyrex [20] and HFSP [42, 43] are a sample of size-based schedulers, which is a family of policies known to drastically improve turnaround times, as we also have verified with our experiments. Similarly, Quincy [17] and DRF [44] study max-min fair, task-level resource allocation, specifically working on multi-dimensional resources. Although our prototype currently consider a one-dimensional packing problem, due to the characteristics of the back-end we use, which does not yet support CPU-level partitioning, ideas presented in [44] can be extended to our work, considering alternative back-ends supporting multi-resource partitioning. Recently, schedulers supporting complex directed acyclic graphs representing low-level, parallel computations have also appeared: Graphene [45], for example, addresses the problem of complex dependencies and heterogeneous demands among the various stages of the computational graph. The work in [46] indicate substantial improvements in terms of resource utilization (and not only allocation) thanks to worker queues, that independently schedule tasks. Bistro [11] employs a novel hierarchical model of data and computational resources, which enable efficient scheduling of data-parallel workloads. Firmament [47] is a centralized scheduler that has been shown to scale to over ten thousand machines at sub-second task placement latency, using a min-cost max-flow optimization approach. Issues related to scheduling scalability, due to the sheer number of low-level tasks that are typically required by analytic jobs, have been addressed through a distributed design, such as in Sparrow [18] and in Condor [12]. Although working at the application-level as we do in our work imposes a low toll on the scheduler, distributed designs are interesting also from the failure tolerance point of view, which is why they represent a valid option for our future work.

# 8 Conclusions

Efficient resource management of computer clusters has been a long-lasting area of research, with peaks of attention happening in conjunction to improvements in computing machinery, e.g. lately with cloud computing and big data. A new breed of cluster management systems, aiming at becoming "data-center operating systems", are currently been confronted with problems of efficiency and performance at scale.

Despite recent advances, there exists a gap between the goal of low-level resource management, and that of manipulating high-level, heterogeneous, distributed (analytic) applications running in such cluster environments. In this paper we presented a first possible step to fill this gap, in the form of a new application scheduler that interacts with a cluster management back-end, to schedule and allocate resources to applications defined with a simple language and semantics. In addition to careful engineering, required us to design and implement our system we call Zoe, our research identified a more fundamental problem, that required to design a novel scheduling heuristic capable of manipulating composite applications, while contributing to system responsiveness.

We validated our algorithm to address our scheduling problem along two lines. We used a numerical approach to simulate large-scale deployments and workloads. We showed our scheduling algorithm to be highly effective in reducing turnaround times, in particular by reducing applications queuing times. Consequently, cluster resources were better allocated. In addition, we reported an overview of the evaluation of Zoe, that indicates superior performance and efficiency related to our flexible scheduling heuristic.

Our road-map includes the development of a method to redeem untapped resources from idle but running applications, which calls for a substantial rethinking of the resource reservation paradigm; the design and implementation of application fault tolerance mechanisms; and a long list of pending “tickets” stemming from our open-source Zoe project.